\documentclass[journal]{IEEEtran}
\usepackage{amsmath,amsfonts, amssymb, amsthm}
\usepackage{algorithmic}
\usepackage{algorithm}
\usepackage{array}
\usepackage{textcomp}
\usepackage{stfloats}
\usepackage{url}
\usepackage{verbatim}
\usepackage{graphicx}
\usepackage{cite}
\usepackage{multicol}
\usepackage{multirow}
\usepackage{colortbl}
\usepackage{bbding}

\usepackage{booktabs,multirow,bm}
\usepackage{tabularx}

\usepackage{xcolor}
\usepackage{lipsum}

\usepackage[table]{xcolor}

\ifCLASSOPTIONcompsoc
  \usepackage[caption=false,font=normalsize,labelfont=sf,textfont=sf]{subfig}
\else
  \usepackage[caption=false,font=footnotesize]{subfig}
\fi

\newcommand{\x}{\boldsymbol{x}}
\newcommand{\s}{\boldsymbol{s}}
\newcommand{\z}{\boldsymbol{z}}
\newcommand{\y}{\boldsymbol{y}}

\newcommand{\h}{\boldsymbol{h}}

\newcommand{\bS}{\boldsymbol{S}}

\newcommand{\bphi}{\boldsymbol{\phi}}

\newcommand{\btheta}{\boldsymbol{\theta}}

\newcommand{\bepsilon}{\boldsymbol{\epsilon}}

\DeclareMathOperator*{\argmin}{argmin}

\newtheorem{remark}{Remark}

\makeatletter

\newcommand{\Rmnum}[1]{\expandafter\@slowromancap\romannumeral #1@}
\makeatother

\newcolumntype{L}[1]{>{\raggedright\arraybackslash}p{#1}}
\begin{document}
	%
    \title{Task-Oriented Communication with Hybrid-Precision Models}
	%
	%
	%
	\author{Songjie~Xie,~\IEEEmembership{Graduate~Student~Member,~IEEE}, Wei~Guo,~\IEEEmembership{Member,~IEEE}, Shenghui~Song,~\IEEEmembership{Senior~Member,~IEEE}, Jun~Zhang,~\IEEEmembership{Fellow,~IEEE}, Ying-Jun~Angela~Zhang,~\IEEEmembership{Fellow,~IEEE}, and Khaled~B.~Letaief,~\IEEEmembership{Fellow,~IEEE}
		\thanks{The authors are with the Department of Electronic and Computer Engineering, The Hong Kong University of Science and Technology, Hong Kong (e-mail: sxieat@connect.ust.hk, \{eeweiguo, eeshsong, eejzhang, eekhaled\}@ust.hk).
        
        Ying-Jun Angela Zhang is with the Department of Information Engineering, The Chinese University of Hong Kong, Hong Kong (email:
yjzhang@ie.cuhk.edu.hk).

}
	}

	\maketitle
	
\begin{abstract} 
Edge inference has emerged as a promising solution for the proliferation of artificial intelligence (AI) services by deploying models at the network edge to circumvent cloud-routing latency. Existing edge inference approaches mainly focused on either cooperative inference to reduce latency or lightweight model design to fit resource-constrained devices. These solutions often address the communication and computation challenges separately, and thus struggle to achieve a balanced trade-off among transmission efficiency, on-device processing cost, and inference accuracy. To bridge this gap, this paper proposes a hybrid-precision task-oriented communication framework for edge inference to holistically balance communication, on-device computation, and utility. In this framework, a binarized front-end is deployed on the edge device to extract and transmit binary features via orthogonal frequency-division multiplexing (OFDM) signals, while a full-precision back-end on the edge server performs the final inference. To ensure model consistency, we introduce an on-device binarization method tailored for split inference and develop an integrated channel-aware transmission scheme featuring subcarrier-based feature calibration. Furthermore, a knowledge distillation (KD)-based training strategy, supported by specialized gradient estimators, is developed to optimize the end-to-end system and inherit semantic knowledge from a full-precision teacher model. Extensive experiments on the large-scale ImageNet dataset demonstrate the superiority of the proposed hybrid system. Our analysis confirms that this design achieves an optimal trade-off among communication efficiency, on-device computational cost, and inference accuracy, outperforming existing edge inference solutions.
\end{abstract}
	

\section{Introduction}
The rapid evolution of artificial intelligence (AI) is fundamentally transforming wireless networks, driving the vision of the future wireless towards supporting ubiquitous intelligent services~\cite{letaief2019roadmap, wang2023road, xie2026towards}. Across a wide spectrum of emerging applications, ranging from virtual/augmented reality (VR/AR) to autonomous driving, the inference based on deep neural networks (DNNs) has emerged as a pivotal enabler for real-time perception and intelligent decision-making~\cite{li2019edge, zhu2020toward}. However, the traditional paradigm of centralized cloud inference, where massive volumes of high-dimensional data are offloaded to remote data centers, suffers from prohibitive communication overhead and high routing latency. To circumvent these systemic bottlenecks and support latency-sensitive tasks, edge inference has emerged as a promising paradigm, where intensive data processing and model execution are strategically shifted from the distant cloud to the network edge~\cite{mao2024green, xu2023edge, letaief2021edge}. 
\begin{figure}[t]
		\centering
		\includegraphics[width=8.7cm]{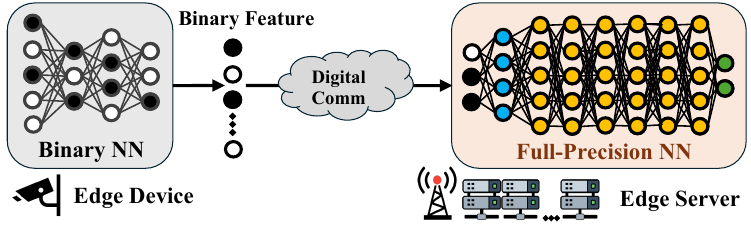}
        \vspace{-5pt}
		\caption{The hybrid-precision framework for edge inference, where the BNN and FPNN are deployed on the edge device and the edge server, respectively, and the binary features are transmitted via digital communication systems. 
		}
        \vspace{-12pt}
        \label{fig: intro}
\end{figure}

Despite its great potential, edge deployment is hampered by the tension between prohibitive data transmission overhead and the resource constraints of edge devices. A straightforward strategy is server-based inference, where edge devices offload raw data to an edge server for full DNN execution~\cite{zhou2019edge}. Nevertheless, offloading heavy computation, transmitting massive raw data results in high communication overhead and raises data privacy concerns~\cite{chen2019deep}. To mitigate these issues, edge-device cooperative inference was proposed by splitting the DNN between the resource-constrained edge device and the edge server, allowing the device to transmit intermediate features instead of raw data~\cite{eshratifar2019jointdnn, shao2020communication}. This cooperative paradigm is further refined by the emerging task-oriented communication, which focuses on extracting and transmitting only task-essential features to reduce the communication overhead and latency~\cite{shao2021learning, gunduz2022beyond, li2026mutual}. This efficiency is further augmented by feature compression and quantization techniques, which map high-dimensional intermediate representations into compact and low-bit discrete values for efficient transmission~\cite{shao2020bottlenet++, xie2023robust}. Nevertheless, from a hardware perspective, the existing split-inference and task-oriented schemes still need to run a full-precision neural network (FPNN) on edge devices, and the required floating-point operations (FLOPs) can remain prohibitive under the stringent compute and power budgets of typical edge devices.

\begin{table*}
\centering
\caption{Edge Inference Method Comparison. }\label{tab: intro}
\vspace{-5pt}
\begin{tabular}{L{2.4cm} L{2.1cm} L{2.1cm} L{2.1cm} L{2.1cm} L{2.1cm} L{2.1cm}}
\toprule
\textbf{Method} & \textbf{Local Data Exposure}  & \textbf{Communication Overhead} & \textbf{Feature Value Type} & \textbf{On-Device FLOPs} & \textbf{On-Device BOPs} & \textbf{Accuracy} \\
\midrule
Server-Based FPNN~\cite{cheng2018model} & \cellcolor{red!20}Yes (Local data exposed to the edge server) & \cellcolor{red!20} High & \cellcolor{green!20} Discrete & \cellcolor{green!20} Minimal & \cellcolor{green!20} Minimal  & \cellcolor{green!20} Very High\\

FPNN-Split~\cite{bourtsoulatze2019deep,gunduz2024joint} & \cellcolor{green!20} No & \cellcolor{red!20} High (High-dimensional features)& \cellcolor{red!20}Continuous &  \cellcolor{red!20} High (Partial FPNN) & \cellcolor{green!20} Low & \cellcolor{green!20} High\\

FPNN-Split-Compression~\cite{shao2020bottlenet++, shao2021learning, shao2020communication} & \cellcolor{green!20} No & \cellcolor{green!20} Low& \cellcolor{red!20}Continuous & \cellcolor{red!20} Very High (Partial FPNN + compression)& \cellcolor{green!20} Low & \cellcolor{green!20} High\\

FPNN-Split-VQ~\cite{hu2023robust,xie2023robust,park2024joint} & \cellcolor{green!20} No & \cellcolor{green!20} Low & \cellcolor{green!20} Discrete (Quantized features) & \cellcolor{red!20} Very High (Partial FPNN + VQ) & \cellcolor{green!20} Low& \cellcolor{green!20} High\\

Full BNN~\cite{rastegari2016xnor, liu2018bi, liu2020reactnet} & \cellcolor{green!20} No (No data transmission) &  \cellcolor{green!20} Minimal (No data transmission) & \cellcolor{green!20} Not Applicable &  \cellcolor{green!20} Low &  \cellcolor{red!20} High (Whole BNN) &  \cellcolor{red!20}Moderate (Reduced due to E2E binarization)\\

\textbf{Proposed Hybrid-Precision} &  \cellcolor{green!20} No &  \cellcolor{green!20} Low & \cellcolor{green!20} Discrete (Binary features) &  \cellcolor{green!20} Low &  \cellcolor{green!20} Low &  \cellcolor{green!20} Very High\\
\bottomrule
\end{tabular}
\vspace{-10pt}
\end{table*}

{To address the on-device hardware bottleneck, a distinct line of research aims to replace computationally expensive FLOPs with highly efficient bitwise operations, named binary neural networks (BNNs)~\cite{rastegari2016xnor, hubara2016binarized}.} Over the years, numerous BNN methods have been developed to binarize AI models end-to-end to enable fully on-device execution~\cite{lin2017towards, qin2020forward, he2020proxybnn, liu2020reactnet, xu2021recu,liu2018bi, tu2022adabin}. However, while avoiding high FLOPs demand, BNNs still incur significant binary operations (BOPs) on the edge device. Moreover, the end-to-end binarization severely restricts model expressivity and introduces substantial quantization errors, resulting in performance degradation that limits their suitability for high-accuracy, mission-critical applications.


Synthesizing these developments, we observe that existing schemes optimize one dimension at the expense of others. Server-based inference minimizes on-device workload but burdens the communication; model-splitting reduces transmission but strains device FLOPs; and end-to-end BNNs reduce FLOPs but sacrifice utility. These limitations underscore that a viable edge inference framework must move beyond fragmented considerations to holistically balance three critical dimensions:
\begin{itemize}
    \item [{\textbf{1)}}] \textbf{Communication:} The system must achieve low communication overhead and ensure compatibility with existing digital communication systems.
    \item [{\textbf{2)}}] \textbf{On-Device Computation:} The inference process needs to adhere to the hardware constraints of edge devices, specifically by minimizing both FLOPs and BOPs to ensure execution efficiency.
    \item [{\textbf{3)}}] \textbf{Utility:} The system must guarantee sufficient inference accuracy, overcoming the performance degradation often associated with quantization or compression.
\end{itemize}
To combine the best of all the worlds, we propose a hybrid-precision task-oriented communication framework for edge inference. As illustrated in Fig.~\ref{fig: intro}, a binarized front-end network is implemented on the edge device, performing the initial processing to extract task-relevant information into binary features. Then, the encoded binary features are transmitted via the digital communication system to the edge server, where the large full-precision back-end network is deployed to perform the final inference. Table~\ref{tab: intro} presents the comparison with the existing methods and highlights the strategic advantages of the proposed method on multiple aspects across communication, on-device computation, and utility. Particularly, the communication overhead is measured by the dimensionality of the transmitted data or features, and the feature value type indicates the compatibility of digital communication systems. The negative for this part is highlighted in red, and otherwise in green.

Despite the expected advantages, developing such a hybrid-precision task-oriented framework navigates several unique technical challenges. The first fundamental challenge is the feature consistency and utility loss of integrating on-device architectural binarization with model splitting. Training a partial BNN on the edge device to generate intermediate binary features risks a severe degradation in overall model accuracy, as these binarized features must retain the essential information required by the full-precision server-side model for the task. Moreover, most existing edge inference studies assume simple channel conditions, such as AWGN or flat fading, neglecting the frequency-selective fading characteristic of realistic wireless environments. For highly compact binary features, the bit corruption caused by the inter-symbol interference can lead to catastrophic utility degradation. To cope with frequency-selective fading, orthogonal frequency division multiplexing (OFDM) needs to be integrated into the task-oriented communication system. However, the non-differentiable nature of OFDM systems together with on-device binarization pose significant challenges to the end-to-end optimization for task-oriented communication systems. 

In this work, we investigate the design of a hybrid-precision edge inference system that seamlessly integrates efficient architectural binarization with co-inference frameworks. To the best of our knowledge, this work introduces the first hybrid-precision solution for edge inference with OFDM systems. To address the presented challenges, including utility loss from feature binarization, reliable feature transmission, and the need for holistic optimization, we develop an efficient and fully integrated framework with holistic optimization. Our core contributions are summarized as follows:
\begin{itemize}
    \item We develop a binarization method for task-oriented wireless edge inference, where the edge device employs a binarized front-end network as a lightweight semantic feature encoder. Unlike conventional BNNs designed for standalone on-device inference, the proposed method is optimized to generate compact binary intermediate features that are both communication-efficient for wireless transmission and semantically compatible with the full-precision back-end deployed at the edge server. 
    \item We design a channel-aware binary feature transmission and recovery mechanism over multipath fading channels. Specifically, we propose a subcarrier-based calibration method that maps modulated binary feature segments to OFDM subcarriers according to channel reliability, enabling task-relevant feature dimensions to be preferentially delivered through high-quality wireless resources. At the edge server, a reliability-aware feature recovery mask suppresses corrupted binary features and reconstructs real-valued representations for subsequent full-precision inference.
    \item We develop a holistic knowledge distillation (KD)-based end-to-end optimization framework for the proposed hybrid-precision task-oriented communication system. The KD-based teacher-student strategy maximizes task-relevant information by transferring knowledge from a full-precision teacher model, while the binary feature representation, subcarrier-based calibration, and wireless-induced feature recovery naturally impose an information bottleneck (IB) on the transmitted features. In this way, the proposed training framework encourages the edge-device encoder to extract compact task-relevant representations while suppressing task-irrelevant redundancy under architectural and wireless transmission constraints. STE-based gradient propagation is further employed to handle the non-differentiability introduced by neural binarization, digital modulation and demodulation, and wireless transmission.
\end{itemize}

 The rest of the paper is organized as follows. Section~\ref{sec: related} presents the related works in edge inference and Section~\ref{sec: model} introduces the system model of the hybrid-precision edge inference system with OFDM. Section~\ref{sec: framework} presents the hybrid-precision edge inference framework, including the design of on-device BNNs, feature calibration, and server-based FPNNs. The KD-based optimization and gradient propagation strategy of the proposed frameworks are presented in Section~\ref{sec: opt}. In Section~\ref{sec: exp}, we provide extensive simulation results to evaluate the performance and effectiveness of the proposed hybrid-precision design. Finally, Section~\ref{sec: conc} concludes the paper.

\section{Related Work} \label{sec: related}
\subsection{Edge Inference}\label{subsec:related-inference}
{Edge inference has shifted from server-based execution to collaborative device-edge intelligence. Early server-centric solutions suffered from high latency and privacy risks when offloading raw data\cite{cheng2018model}. While edge co-inference mitigates this by partitioning DNNs between devices and servers, it faces the \emph{data amplification} issue, where intermediate features exceed the original input size~\cite{shao2020communication}. Thus, specialized compression is vital to reduce dimensionality and achieve low-latency inference~\cite{shao2020bottlenet++}.}

Concurrently, the advancement of learning-based communication has redefined feature transmission strategies. Joint source-channel coding (JSCC) utilizes DNN-based encoders and decoders to transmit information over wireless channels, showing significant success in image and text transmission under various channel models~\cite{gunduz2024joint,yang2022ofdm,bourtsoulatze2019deep}. Building on this, the task-oriented communication paradigm was developed to shift the objective from data reconstruction to successful task completion~\cite{shi2023task}. By leveraging the information bottleneck (IB) principle, these frameworks discard task-irrelevant information, significantly enhancing communication efficiency~\cite{shao2021learning}. Recent studies have further extended this principle to address multi-device cooperation~\cite{shao2022task, cai2024multi}, out-of-distribution (OOD) robustness through invariant risk minimization~\cite{li2024tackling}, and cross-model alignment for heterogeneous edge environments~\cite{xie2025toward}.

Despite the efficiency of task-oriented schemes, the reliance on continuous floating-point feature values poses significant integration challenges for modern digital radio frequency (RF) systems~\cite{park2024joint}. This has motivated the incorporation of discrete representation learning techniques, such as vector quantization (VQ)~\cite{xie2022task}. Frameworks utilizing VQ-based IB or vector quantized-variational autoencoders (VQ-VAEs) enable digital transmission by mapping features into discrete codebooks, often coupled with adaptive modulation to enhance robustness against channel noise~\cite{xie2023robust}.

However, existing model-splitting and discrete representation solutions frequently overlook the hardware-level constraints of edge devices, particularly regarding FLOPs capacity. The deployment of standard DNN layers, alongside additional compression and VQ modules, often exceeds the computational budget of low-power edge hardware. This computational bottleneck remains a critical barrier to the practical deployment of sophisticated edge AI models, necessitating the exploration of ultra-low-precision architectures like the one proposed in this work.

\subsection{Binary Neural Network}\label{subsec:related-BNN}
{The fundamental motivation for BNNs in edge intelligence is to replace resource-intensive floating-point operations with bitwise XNOR and popcount operations. However, the non-differentiable nature of binarization and the inherent loss of information represent significant hurdles. Existing research to address these challenges can be categorized into two main trajectories: precision-centric optimization and architecture and deployment efficiency.

Early research focused on minimizing the quantization error between full-precision and binary representations. XNOR-Net~\cite{rastegari2016xnor} introduced analytic scaling factors to close the accuracy gap, while ABC-Net~\cite{lin2017towards} utilized multiple binary bases to approximate full-precision weights. To stabilize the training process and mitigate gradient mismatch caused by the sign function, IR-Net~\cite{qin2020forward} and ProxyBNN~\cite{he2020proxybnn} proposed information-retention and proxy-function methods. Recent advances like ReActNet~\cite{liu2020reactnet} and ReCU~\cite{xu2021recu} further refined this by introducing generalized activation functions and rectified clamping to preserve distribution entropy. Despite these innovations, a persistent performance gap remains compared to full-precision counterparts.

Parallel to training optimizations, architectural innovations like Bi-Real-Net~\cite{liu2018bi} introduced shortcut connections to preserve signal flow. This evolved into automated discovery via Neural Architecture Search (NAS), with frameworks like BNAS~\cite{ding2021bnas} finding that wider layers and specific skip-connections are more resilient to binarization noise. To further reduce the overhead of deployment, DIR-Net~\cite{qin2023distribution} and AdaBin~\cite{tu2022adabin} focused on discriminative features and adaptive quantization levels to ensure the network remains lightweight yet task-effective.

Despite these advances, BNNs are typically treated as monolithic entities for on-device deployment. They still encounter the \emph{resource-accuracy} paradox: end-to-end BNNs often lack the precision for complex tasks, yet scaling them to improve accuracy significantly increases on-device BOPs. Furthermore, existing BNNs are not optimized for the split-inference paradigm, where binary features must be both computationally efficient and robust enough to serve as descriptors for a high-precision server-side back-end.}

\begin{figure*}[t]
		\centering
		\includegraphics[width=17cm]{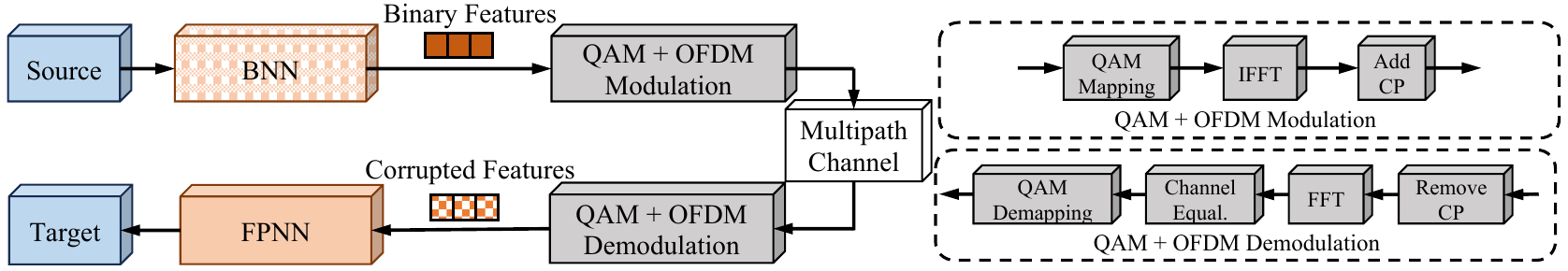}
        \vspace{-5pt}
		\caption{The system model of the hybrid edge inference system with OFDM.
		}
        \vspace{-10pt}
  \label{fig:system_model}
\end{figure*}
\section{System Model}\label{sec: model}
As shown in Fig.~\ref{fig:system_model}, we consider a point-to-point OFDM-based task-oriented communication system for edge inference. Specifically, the edge device utilizes a BNN for feature encoding and outputs the binary features. The binary features are represented by the quadrature amplitude modulation (QAM) and fed to the OFDM extension to generate OFDM symbols. After transmission over the multipath channels, the edge server decodes the received signals using the QAM and OFDM demodulator and performs the inference based on the FPNN. The details are stated as follows.

\subsection{The BNN-Based Transmitter with OFDM} At the edge device, there is a data source $\x \in \mathcal{X}$ associated with its target $\y \in \mathcal{Y}$ (e.g., the label for the image) with an underlying joint distribution $p(\x, \y)$. The BNN $f_{\bar\btheta}$ served as a feature encoder deployed at the edge device with the binarized parameters $\bar \btheta$. The data $\x$ is encoded by $f_{\bar\btheta}$ into the bitstream
\begin{align}
    \boldsymbol{b} & = f_{\bar\btheta}(\x),
\end{align}
where $\boldsymbol{b}\in \{0,1 \}^{K_b}$ denote the encoded binary features and $K_b$ is the length of the encoded bitstream. Then, the transmitter modulates encoded bitstream $\boldsymbol{b}$ with fixed-size constellations (e.g., 16-QAM) by
\begin{align}
    \boldsymbol{m} &= \mathcal{M}(\boldsymbol{b}),
\end{align}
where $\boldsymbol{m} = (m_1, m_2, \dots, m_{K_m}) \in \mathbb{C}^{K_m}$ denotes the modulated complex signals, $\mathcal{M}(\cdot)$ stands for the fixed-size modulation, and $K_m$ is the length of modulated signals. The relation between the dimensions of $\boldsymbol{m}$ and $\boldsymbol{b}$ is determined by the number of bits represented by the constellations in $\mathcal{M}$, such as $K_m = K_b/4$ for 16-QAM. Then, we need to reshape $\boldsymbol{m}$ into $N_s$ OFDM symbols,
\begin{align}
    \bS & = (\s_1, \s_2, \dots, \s_{N_s}) \in\mathbb{C}^{N_s\times N_{\rm{FFT}}},
\end{align}
where $\bS$ is the collection of $N_s$ OFDM symbols and $N_{\rm{FFT}}$ is the number of subcarriers. Specifically, the reshaping operation ensures $K_m = N_s \times N_{\rm{FFT}}$.

Subsequently, the inverse fast Fourier transform (IFFT) is applied to each symbol, and the cyclic prefix (CP) with the length of $N_{CP}$ is inserted into the OFDM
signals to ensure subcarrier orthogonality. After that, the transmitter transmits the time domain signals $\z \in \mathbb{C}^{N_s \times (N_{\rm{FFT}}+N_{CP})}$ propagates through the multipath fading channel.

\subsection{Signal Propagation over Wireless Multipath Channel} 
We consider a $\Psi$-tap discrete-time multipath fading channel. The received time-domain signal is modeled as the convolution of the transmitted OFDM signal with the channel impulse response (CIR) plus additive noise, i.e.,
\begin{align}
    \tilde{\boldsymbol{z}} = \boldsymbol{\eta} * \boldsymbol{z} + \boldsymbol{\epsilon},
\end{align}
where $*$ denotes linear convolution, $\boldsymbol{\eta} \in \mathbb{C}^\Psi$ denotes the CIR with $\Psi$ propagation paths, each experiences independent Rayleigh fading satisfying {$\eta_{\psi} \sim \mathcal{CN}(0, \sigma_{\psi}^2)$ for $\psi= 1, 2, \dots, \Psi$}, and $\bepsilon \sim \mathcal{CN}(\mathbf{0}, \sigma^2 \boldsymbol{I})$ is the additive complex Gaussian noise.  The power of each path follows an exponential decay profile $\sigma^2_{\psi} = \alpha_{\psi}\exp(-\frac{\psi}{\gamma})$, where $\alpha_{\psi}$ is a normalization coefficient to satisfy $\sum_{\psi=1}^{\Psi} \sigma_{\psi}^2 = 1$ and $\gamma$ is the delay spread constant.

For the transmitted OFDM signals, the channel response in the frequency domain for the subcarriers is denoted as $\h = (h_1, h_2, \dots, h_{N_{\rm{FFT}}})\in \mathbb{C}^{N_{\rm{FFT}}}$. Then, for the $i$-th OFDM symbol, the received frequency-domain symbols $\tilde{\s}_i = (\tilde{s}_{i,1}, \tilde{s}_{i,2}, \dots, \tilde{s}_{i, N_{\rm{FFT}}})$ satisfy
\begin{align}
    \tilde{s}_{i,k} &= h_ks_{i,k} + n_{i,k}, \quad k \in \{1, 2, \dots, N_{\rm{FFT}}\},
\end{align}
where $n_{i,k}$ denotes the noise on the $k$-th subcarrier of the $i$-th received OFDM symbol.

\subsection{The FPNN-based Receiver with OFDM extension} At the edge server, the receiver first removes the CP from the received time-domain signal $\tilde{\s}$, and then transforms it into frequency-domain signals by applying fast Fourier transform (FFT), which denoted as $\tilde{\bS} = (\tilde{\s}_1, \tilde{\s}_2, \dots, \tilde{\s}_{N_s})\in \mathbb{C}^{N_s\times N_{\rm{FFT}}}$. We assume that each sub-carrier has a bandwidth that is much smaller than the coherence bandwidth of the channel. The instantaneous channel estimations on all the subcarriers can be estimated at the receiver, denoted as $\hat{\h} = (\hat{h}_1, \hat{h}_2, \dots, \hat{h}_{N_{\rm{FFT}}})$. Consequently, conventional equalizers such as the zero-forcing (ZF) or minimum mean square error (MMSE) equalizers can be adopted to efficiently mitigate frequency-selective fading. For instance, the OFDM signal obtained through simple frequency-domain equalization is shown as,
\begin{align}
    \hat{s}_{i,k} & = \frac{\tilde{s}_{i,k}}{\hat{h}_k} = \frac{h_k}{\hat{h}_k}s_{i,k} + \frac{n_{i,k}}{{\hat{h}_k}},
\end{align}
where $\hat{s}_{i,k}$ is the $k$-th dimension of the equalized symbol $\hat{\s}_i = (\hat{s}_{i,1}, \hat{s}_{i,2}, \dots, \hat{s}_{i, N_{\rm{FFT}}})$. The equalized signals $\hat{\bS}= (\hat{\s}_1, \hat{\s}_2, \dots, \hat{\s}_{N_s})$ will be reshaped into the sequence of complex signals $\Tilde{\boldsymbol{m}} \in \mathbb{C}^{K_m}$ and then $\Tilde{\boldsymbol{m}}$ is demodulated into the bitstream $\Tilde{\boldsymbol{b}}$ with fixed-size constellations by 
\begin{align}
    \tilde{\boldsymbol{b}} & = \mathcal{M}^{-1}(\tilde{\boldsymbol{m}}).
\end{align}
The obtained bitstream, treated as binary features, is input into the FPNN $f_{\bphi}$ deployed at the edge server with full-precision neural network parameters $\bphi$. Then, the FPNN outputs the prediction of the target $\hat{\z}$ by $\hat{\z} = f_{\bphi}(\tilde{\boldsymbol{b}})$.


\section{Hybrid Precision Edge Inference Framework}\label{sec: framework}
\begin{figure*}[t]
		\centering
		\includegraphics[width=18cm]{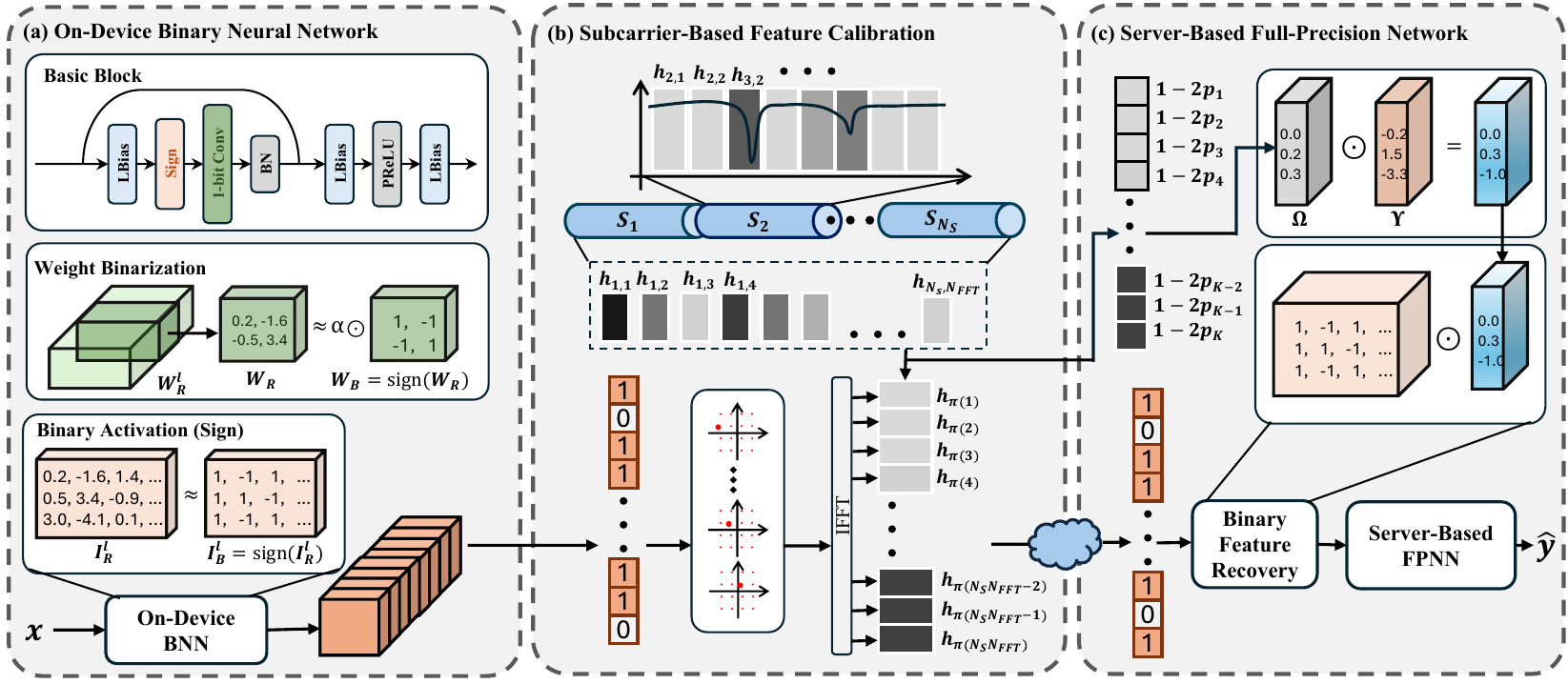}
        \vspace{-5pt}
		\caption{The proposed hybrid-precision task-oriented communication system for edge inference.
		}\label{fig: framework}
        \vspace{-10pt}
\end{figure*}

In this section, we present the hybrid-precision task-oriented communication system by developing the on-device binarization and server-based FPNNs. Furthermore, the subcarrier-based feature calibration and binary feature recovery are developed to enhance the binary feature transmission via OFDM. The overall design of the proposed edge inference framework is illustrated in Fig.~\ref{fig: framework}.
\subsection{On-Device BNN}\label{subsec: BNN} 
To facilitate low-latency execution on resource-constrained hardware, we employ an $L$-layer binary convolutional neural network (CNN) on the edge device. For the $l$-th layer,  let $\mathbf{I}^l \in \mathbb{R}^{C_l\times H_l \times W_l}$ denote the input activation tensor and $\mathbf{W}^l \in \mathbb{R}^{C_{l+1}\times C_l \times K \times K}$ represent the weight filters. 
To maintain clarity in our formulation, we distinguish between full-precision and binary representations using subscripts: real-valued tensors are denoted as $\mathbf{I}^l_{\rm{R}}$ and $\mathbf{W}^l_{\rm{R}}$, while their corresponding binarized counterparts are denoted as $\mathbf{I}^l_{\rm{B}}$ and $\mathbf{W}^l_{\rm{B}}$, respectively\footnote{The real-valued tensors $\mathbf{I}^l_{\rm{R}}$ and $\mathbf{W}^l_{\rm{R}}$ are latent parameters used exclusively during the training phase for gradient updates. Upon deployment, the model executes inference using only the binary parameters ($\mathbf{I}^l_{\rm{B}}$, $\mathbf{W}^l_{\rm{B}}$), thereby incurring no additional storage or computational overhead.}.

\subsubsection{Weight Binarization} 
To preserve model capacity while transitioning to discrete weights, we approximate each real-valued filter $\mathbf{W}_{\rm{R}} \in \mathbf{W}^l_{\rm{R}}$ using a binary filter $\mathbf{W}_{\rm{B}} \in \mathbf{W}^l_{\rm{B}}$ scaled by a positive factor $\alpha \in \mathbb{R}^+$. The objective is to minimize the reconstruction error such that $\mathbf{W}_{\rm{R}} \approx \alpha \mathbf{W}_{\rm{B}}$. Formally, we seek the optimal $\mathbf{W}_{\rm{B}}^*$ and $\alpha^*$ that minimize the following least-squares objective:
\begin{align}
    \mathbf{W}_{\rm{B}}^*, \alpha^* = \argmin\limits_{\mathbf{W}_{\rm{B}}, \alpha} \| \mathbf{W}_{\rm{R}} - \alpha \mathbf{W}_{\rm{B}}\|^2.\label{eq: binary-weight}
\end{align}
Following the pioneer work of XNOR-Net~\cite{rastegari2016xnor}, by expanding the least squared objective of \eqref{eq: binary-weight}, we obtain,
\begin{align}
    \alpha^2 \mathbf{W}_{\rm{B}}^T \mathbf{W}_{\rm{B}} -2\alpha\mathbf{W}_{\rm{R}}^T\mathbf{W}_{\rm{B}} + \mathbf{W}_{\rm{R}}^T\mathbf{W}_{\rm{R}}.
\end{align}
Since $\mathbf{W}_{\rm{B}} \in \{+1, -1\}^n$, we have $\mathbf{W}_{\rm{B}}^T \mathbf{W}_{\rm{B}} = n$, where $n$ is the number of elements in the filter. Given that $\mathbf{W}_{\rm{R}}^T \mathbf{W}_{\rm{R}}$ is constant for a fixed real-valued weight, the optimization objective simplifies to $\alpha^2 n-2\alpha\mathbf{W}_{\rm{R}}^T\mathbf{W}_{\rm{B}} + \text{const}$. Therefore, the optimal $\mathbf{W}_{\rm{B}}^*$ is achieved by taking the element-wise sign of the real-valued tensor $\mathbf{W}_{\rm{R}}$:
\begin{align}
    \mathbf{W}^*_{\rm{B},i} & = \text{sign}(\mathbf{W}_{\rm{R}})_i = \left\{ \begin{array}{lr}
         +1 &\mathbf{W}_{\rm{R},i} \geq 0 \\
         -1 &\mathbf{W}_{\rm{R},i} < 0
    \end{array} \label{eq: binary weight}
    \right..
\end{align}
By substituting $\mathbf{W}_{\rm{B}}^* = \text{sign}(\mathbf{W}_{\rm{R}})$ back into the objective function, the optimal scaling factor $\alpha^*$ is derived as the mean of the absolute values of the real-valued filter:
\begin{align}
    \alpha^* &= \frac{\mathbf{W}_{\rm{R}}^T\mathbf{W}_{\rm{B}}^*}{n} = \frac{\| \mathbf{W}_{\rm{R}} \|_1}{n},
\end{align}
where $\| \cdot \|_1$ denotes the $\ell_1$-norm.

While weight binarization significantly reduces computational complexity, achieving the full computational benefits of BNNs requires substituting expensive floating-point convolutions with efficient bitwise operations. This necessitates the binarization of input activations at each layer through a binary activation function.
\subsubsection{Binary Activation}The binary activation $\mathbf{I}_{\rm{B}}$ is obtained by applying the element-wise sign function to the real-valued features $\mathbf{I}_{\rm{R}}$, such that $\mathbf{I}_{\rm{B}} = \text{sign}(\mathbf{I}_{\rm{R}})$. {Note that unlike weight binarization, the magnitude of the activation tensor $\mathbf{I}_{\rm{R}}^l$ is intentionally discarded to further reduce FLOPs.}  By combining the binary weights $\mathbf{W}_{\rm{B}}^{l} = \text{sign}(\mathbf{W}_{\rm{B}}^{l})$ and binary activations $\mathbf{I}_{\rm{B}}^l$, the computationally intensive floating-point convolution in the $l$-th layer can be approximated using highly efficient bitwise operations:

\begin{align}
    \mathbf{W}_{\rm{R}}^l * \mathbf{I}_{\rm{R}}^l & \approx \alpha^* ( \text{sign}(\mathbf{W}_{\rm{R}}^{l}) \circledast \mathbf{I}_{\rm{B}}^l),
\end{align}
where $*$ denotes the standard real-value convolution operation and $\circledast$ represents the binary convolution, which can be implemented via efficient bitwise logic, such as $\text{XNOR}$ and $\text{popcount}$~\cite{rastegari2016xnor}. This substitution of floating-point multiply-accumulate (MAC) operations significantly reduces the power consumption and memory footprint of the edge device. 

\subsubsection{Basic Architectural Block} 
However, static scaling factors often fail to accommodate the varying distributions of binarized input tensors, leading to significant information loss. To dynamically adapt to these distributions, we introduce a trainable bias to the input tensors prior to binarization. This allows the network to learn an optimal activation threshold during training. Inspired by the learnable thresholds in ReActNet~\cite{liu2020reactnet}, we propose an \textit{elastic binarization} function characterized by the learnable parameter $\beta_0 \in \mathbb{R}$:
\begin{align}
    \mathbf{I}_{\rm{B}}^l & = \text{sign}(\mathbf{I}_{\rm{R}}^l + \beta_0).
\end{align}
To further mitigate the fidelity loss induced by binarization, we incorporate a residual shortcut for real-valued features that bypasses the binary convolution. Additionally, learnable biases $\beta_1$ and $\beta_2$ are introduced following the binary convolution and the shortcut summation, respectively. These parameters serve to shift and scale the feature distribution, optimizing the activation range of the PReLU function. Consequently, the output features of the $l$-th layer are expressed as:
\begin{align}
    \mathbf{I}_{\rm{R}}^{l+1} & = \text{PReLU}(\alpha^* ( \text{sign}(\mathbf{W}_{\rm{R}}^{l}) \circledast \mathbf{I}_{\rm{B}}^l) + \beta_1) + \beta_2 + \mathbf{I}_{\rm{R}}^l.
\end{align}

\subsubsection{Feature Binarization before Transmission} 
Following the $L$-layer binarized front-end, the input $\boldsymbol{x}$ is transformed into the output feature map $\mathbf{I}_{\rm{R}}^{L+1} \in \mathbb{R}^{C_L \times H_L \times W_L}$. Crucially, while the internal weights and activations of the front-end are binarized, the inclusion of residual shortcuts results in these final features remaining real-valued. To facilitate transmission over a digital communication, these real-valued features $\mathbf{I}_{\rm{R}}^{L+1}$ must be decomposed into a binary bit sequence according to our proposed encoding mechanism. To preserve the semantic integrity of the features and keep model consistency during this binarization, we adopt scaling factors $\boldsymbol{\Upsilon} \in \mathbb{R}^{+, W_L\times H_L}$ and perform spatial-wise scaling for the binarization. Specifically, for each spatial coordinate $(i, j)$, where $0 < i \leq H_L$ and $0 < j \leq W_L$, we approximate the feature vector $\mathbf{I}_{\rm{R}}^{L+1}[:, i, j]$ using a binary vector $\mathbf{I}_{\rm{B}}^{L+1}[:, i, j] \in \{-1, 1\}^{C_L}$ and a local scaling factor ${\Upsilon}[i,j] \in \mathbb{R}^+$. We formulate this as a reconstruction optimization problem to minimize the binarization error:
\begin{align}
    \mathbf{I}_{\rm{B}}^{L+1}[:, i, j]^*, {\Upsilon} [i,j]^* & = \notag\\ \argmin\limits_{\mathbf{I}_{\rm{B}}^{L+1}[:, i, j], {\Upsilon} [i,j]} \| {\Upsilon} [i,j]&\mathbf{I}_{\rm{B}}^{L+1}[:, i, j] - \mathbf{I}_{\rm{R}}^{L+1}[:, i, j]\|^2.
\end{align}
Following a derivation similar to the weight binarization in \eqref{eq: binary-weight}, the optimal binary features and scaling factors are analytically determined by:
\begin{align}
    \mathbf{I}_{\rm{B}}^{L+1}[:, i, j]^* &= \text{sign}(\mathbf{I}_{\rm{R}}^{L+1}[:, i, j])\\
    {\Upsilon} [i,j] &= \frac{\| \mathbf{I}_{\rm{R}}^{L+1}[:, i, j]\|_1}{C_L}. \label{eq: scaling_transmission}
\end{align}
To finalize the transmission payload, the scaling factors $\boldsymbol{\Upsilon}$ are quantized and concatenated with the binary feature maps $\mathbf{I}_{\rm{B}}^{L+1}$. This combined representation is reshaped into a unified bit sequence $\mathbf{b} \in \{0, 1\}^{K_b}$. Finally, this sequence is mapped onto complex signal constellations using $M$-QAM modulation, enabling efficient transmission over the physical layer.
\subsection{Subcarrier-Based Feature Calibration} 
To ensure robustness in frequency-selective fading environments, we introduce a subcarrier-based feature calibration scheme that aligns feature transmission with subcarrier reliability. In each transmission block, there are $N_{\rm{FFT}}$ subcarriers for each OFDM symbol, and each data point $\x$ requires $N_s$ symbols to fully transmit the binary features $\boldsymbol{b}$. Each subcarrier is identified by a pair $(t, k)$, where $t=1, 2, \dots, N_s$ is the OFDM symbol index, and $k=1,2, \dots, N_{\rm{FFT}}$ is the subcarrier index within the symbol. Furthermore, the subcarrier $(t,k)$ corresponds to the OFDM signal in the OFDM symbols $s_{t,k} \in \mathbf{S}$. The quality of subcarrier $(t, k)$ is represented as its channel gain $|h_{t,k}|^2$, where $|\cdot|^2$ denotes the squared magnitude of the complex channel coefficient. We sort these subcarriers in descending order of their channel gains to obtain a reliability permutation $\pi$, such that
\begin{align}
    |h_{\pi(1)}|^2 \geq |h_{\pi(2)}|^2 \geq \dots \geq |h_{\pi(N_s N_{\text{FFT}})}|^2,
\end{align}
where $\pi(m) = (t_m, k_m)$ maps the $m$-th best subcarrier to its corresponding OFDM symbol $t_m$ and the subcarrier index $k_m$. 

Rather than utilizing a random mapping, which would subject critical semantic bits to unpredictable channel fades, our scheme establishes a deterministic hierarchy of transmission slots. We consecutively map the modulated feature segments $\mathbf{M} = (\boldsymbol{m}_1, \dots, \boldsymbol{m}_{K_m})$ to the ordered subcarriers:
\begin{align}m_i \to s_{\pi(i)}, \quad \text{for } i = 1, \dots, N_s N_{\text{FFT}}.\end{align}
By fixing these \emph{safe positions} according to channel reliability, the framework enables the model to implicitly learn during training which feature dimensions are most critical and should be encoded into these prioritized slots. As illustrated in Fig.~\ref{fig: framework}, this ordered mapping ensures that the most essential task-relevant information is consistently protected by the highest-quality channel resources, effectively maximizing inference utility in challenging multipath conditions\footnote{Furthermore, the quantized bits of the spatial scaling factors $\boldsymbol{\Upsilon}$ are prioritized for transmission over the highest-quality subcarriers. This ensures that these critical values remain error-free, providing a reliable foundation for the subsequent feature recovery process.}. 

\subsection{Server-Based FPNN}
Following OFDM transmission over the multipath fading channel, the received bitstream $\tilde{\boldsymbol{b}}$ is reshaped into binary feature maps $\tilde{\mathbf{I}}^{L+1}_{\rm{B}}\in \{-1, +1\}^{C_L \times H_L \times W_L}$ and provided to the server-side FPNN alongside the scaling factors $\boldsymbol{\Upsilon}$. However, directly performing inference on features corrupted by wireless noise leads to a catastrophic degradation in task utility. To mitigate this, we adopt a real-valued tensor $\boldsymbol{\Omega} \in \mathbb{R}^{C_L \times H_L \times W_L}$ as a mask and apply it on $\tilde{\mathbf{I}}^{L+1}_{\rm{B}}$ to filter bit errors while mapping the received bits back into the continuous feature domain. 

First, to characterize the impairments introduced during transmission, each element of the transmitted feature tensor $\mathbf{I}^{L+1}_{\rm{B}}$ is modeled as passing through an equivalent binary symmetric channel (BSC). Specifically, for a given spatial coordinate $(i, j)$, where $0 \leq i < H_L$ and $0 \leq j < W_L$, the bit errors are abstracted as a vector of independent Bernoulli variables $\mathbf{N}[:, i, j] = (N_1, N_2, \dots, N_{C_L})^T$. Each variable $N_k$ follows the distribution:
\begin{align}
    \Pr[N_k = a] &=\left\{ \begin{array}{lr}
         p_k & a = -1 \\
         1-p_k & a= +1
    \end{array}
    \right.,
\end{align} where $0.5 \geq p_k \geq 0$ is the error probability of the BSC, which can be estimated as the bit error rate of QAM modulation under the corresponding subcarrier channel gain and noise level. Therefore, the corrupted binary feature vector $\Tilde{\mathbf{I}}_{\rm{B}}^{L+1}[:, i, j]$ can be expressed as $\Tilde{\mathbf{I}}^{L+1}_{\rm{B}}[:, i, j] = \mathbf{N}[:, i, j] \odot \mathbf{I}_{\rm{B}}^{L+1}[:, i,j]$, where $\odot$ denotes the element-wise multiplication. 

For notational clarity and owing to the spatial symmetry of the feature maps, we henceforth omit the spatial indices and utilize $\mathbf{I}_{\rm{B}}$, $\mathbf{I}_{\rm{R}}$, $\mathbf{n}$, $\boldsymbol{\omega}$ and $\upsilon$ to represent $\mathbf{I}_{\rm{B}}^{L+1}[:, i, j]$, $\mathbf{I}_{\rm{R}}^{L+1}[:, i, j]$, $\mathbf{N}[:, i, j]$, $\boldsymbol{\Omega}[:, i, j]$ and $\boldsymbol{\Upsilon}[i, j]$, respectively.  

Similar to the binarization process, we aim to minimize the information loss between $\upsilon \cdot \boldsymbol{\omega}\odot \mathbf{n} \odot \mathbf{I}_{\rm{B}}$ and $\mathbf{I}_{\rm{R}}$, which can be formulated as the following stochastic optimization problem
\begin{align}
   \boldsymbol{\omega}^* = \argmin\limits_{\boldsymbol{\omega}} \mathbb{E}[\|\upsilon \cdot \boldsymbol{\omega}\odot \mathbf{n} \odot \mathbf{I}_{\rm{B}} - \mathbf{I}_{\rm{R}} \|^2]. \label{eq:noise-obj}
\end{align}
By expanding the objective function in \eqref{eq:noise-obj}, we have
\begin{align}
    &\mathbb{E}[\upsilon^2(\boldsymbol{\omega}\odot \mathbf{n} \odot \mathbf{I}_{\rm{B}})^T(\boldsymbol{\omega}\odot \mathbf{n} \odot \mathbf{I}_{\rm{B}})] \notag\\
    &\quad \quad \quad \quad - 2\upsilon \mathbb{E}[(\boldsymbol{\omega}\odot \mathbf{n} \odot \mathbf{I}_{\rm{B}})^T\mathbf{I}_{\rm{R}}] + \mathbf{I}_{\rm{R}}^T\mathbf{I}_{\rm{R}}\notag\\
    =& \upsilon^2 \sum\limits^{C_L}_{k=1}\mathbb{E}[\boldsymbol{\omega}_k^2\mathbf{n}_k^2 \mathbf{I}_{\rm{B},k}^2] -2 \upsilon \sum\limits_{k=1}^{C_L} \mathbb{E}[\boldsymbol{\omega}_k\mathbf{n}_k \mathbf{I}_{\rm{B},k} \mathbf{I}_{\rm{R},k}] + \sum\limits_{k=1}^{C_L} \mathbf{I}_{\rm{R},k}^2\notag\\
    =& \upsilon^2 \sum\limits^{C_L}_{k=1}\boldsymbol{\omega}^2_k\mathbf{I}_{\rm{B},k}^2 -2 \upsilon \sum\limits_{k=1}^{C_L} (1-2p_k)\boldsymbol{\omega}_k\mathbf{I}_{\rm{B},k} \mathbf{I}_{\rm{R},k}+ \sum\limits_{k=1}^{C_L} \mathbf{I}_{\rm{R},k}^2, \label{eq: random-last}
\end{align}
where the last equivalence is obtained by taking the first and second order moments of the Bernoulli variable $N_k$
\begin{align}
    \mathbb{E}[N_k] & = 1- 2p_k\\
    \mathbb{E}[N^2_k] & = (-1)^2p_k + 1^2(1-p_k)=1.
\end{align}
\begin{figure}[t]
		\centering
		\includegraphics[width=6.cm]{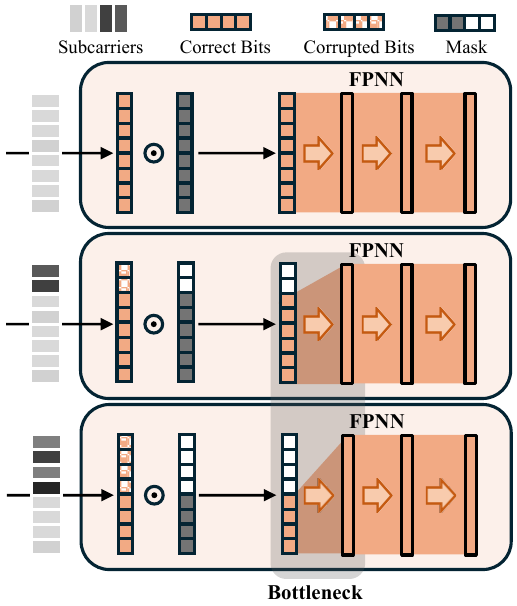}
        \vspace{-5pt}
		\caption{The feature bottleneck induced by the proposed feature calibration and binary feature recovery. When the channel condition is getting worse and more features are corrupted, more dimensionality of transmitted features is masked to reduce the impact of channel noise and reduce the information complexity of received features.
		}
        \vspace{-15pt}
  \label{fig:bottleneck}
\end{figure}
By taking the derivative of \eqref{eq: random-last}, the optimal scaling factor can also be obtained as
\begin{align}
    \boldsymbol{\omega}_k^* =\frac{(1-2p_k)\mathbf{I}_{\rm{B},k} \mathbf{I}_{\rm{R},k}}{\upsilon}. 
\end{align}
As shown in \eqref{eq: binary weight} and \eqref{eq: scaling_transmission}, we have $\mathbf{I}_{\rm{B},k} = \text{sign}(\mathbf{I}_{\rm{R},k})$ and $\upsilon = \| \mathbf{I}_{\rm{R}}\|_1/C_L$ , and we can obtain
\begin{align}
    \boldsymbol{\omega}_k^* &=\frac{(1-2p_k)\text{sign}(\mathbf{I}_{\rm{R},k}) \mathbf{I}_{\rm{R},k}}{ \| \mathbf{I}_{\rm{R}}\|_1/C_L} \\
    &= (1-2p_k)\frac{C_L|\mathbf{I}_{\rm{R},k}|}{\sum^{C_L}_{k=1}|\mathbf{I}_{\rm{R},k}|}.
\end{align}
{As derived, the optimal mask $\boldsymbol{\omega}_k^*$ comprises two components: the channel reliability factor $(1-2p_k)$ and the feature magnitude ratio $\frac{C_L|\mathbf{I}_{\rm{R},k}|}{\sum^{C_L}_{j=1}|\mathbf{I}_{\rm{R},j}|}$. However, as transmitting the individual feature magnitudes $|\mathbf{I}_{\rm{R},k}|$ would incur prohibitive communication overhead, this is strictly unavailable at the receiver. To address this, we approximate the optimal mask by assuming a uniform magnitude distribution for each spatial position, which reduces the magnitude ratio to $1$. Therefore, in our framework, we adopt $\boldsymbol{\omega}_k^* = 1 - 2p_k$ and the mask $\boldsymbol{\Omega}[:, i, j] = \boldsymbol{\omega}^*$ for each spatial coordinate $(i,j)$. The recovered real-valued features $\hat{\mathbf{I}}_{\rm{R}}^{L+1}$ are reconstructed as:}
\begin{align}
    \hat{\mathbf{I}}_{\rm{R}}^{L+1} &=   \boldsymbol{\Upsilon}\odot \boldsymbol{\Omega} \odot\mathbf{\Tilde{I}}_{\rm{B}}^{L+1},\label{eq: recover}
\end{align}
where the spatial scaling factors $\boldsymbol{\Upsilon} \in \mathbb{R}^{H_L \times W_L}$ is broadcast across the channel dimension $C_L$ such that $ \hat{{I}}_{\rm{R}}^{L+1}[k, i,j] =\Upsilon [i,j] \Omega[k, i,j] \Tilde{I}_{\rm{B}}^{L+1}[k, i,j]$. Then, the server-based FPNN leverage $\hat{\mathbf{I}}_{\rm{R}}^{L+1}$ to perform the inference.
\begin{remark}
    The binary feature recovery \eqref{eq: recover} can be interpreted as a noise filter on $\mathbf{\Tilde{I}_B}^{l+1}$ based on the quality of subcarriers. To be specific, if the quality of the subcarrier is really bad, the error probability $p_k \to 0.5$ then $\boldsymbol{\Omega}_k \to 0$. If the quality of the subcarrier is really good, the $p_k \to 0$ then $\boldsymbol{\Omega}_k \to 1$. Because of the calibration and the mask $\boldsymbol{\Omega}$ of binary feature recovery, it will induce a bottleneck, as shown in Fig.~\ref{fig:bottleneck}. The induced bottleneck not only contributes to controlling the uncertainty of noisy feature processing but also enhances the training of the feature encoding that will be discussed in the next section.
\end{remark}

\section{The End-To-End Training for Hybrid Systems}\label{sec: opt}
In this section, we develop the KD-based objective function and the model training strategy for the hybrid-precision task-oriented communication system. 
\subsection{KD-Based Optimization for Hybrid Frameworks}
Training the hybrid-precision architecture presents significant challenges due to the restricted expressivity of the on-device binarized neural parameters. Such binarized architectures often lack the flexibility required to capture the complex data distributions only implied by sparse categorical labels, leading to suboptimal convergence when trained solely with standard cross-entropy (CE) loss. Therefore, we employ knowledge distillation (KD) to bridge this expressivity gap and provide richer supervisory signals. Specifically, a pre-trained FPNN is adopted as a teacher model, $p_{\boldsymbol{T}}(\boldsymbol{y}|\boldsymbol{x})$, to guide the hybrid model by minimizing the Kullback-Leibler (KL) divergence between their respective outputs. Consequently, the holistic training objective $\mathcal{L}$ is formulated as a combination of the CE loss with ground-truth labels and the KD loss:
\begin{align}
    \mathcal{L} &= \mathbb{E}_{p(\tilde{\boldsymbol{b}}|\x) p(\x, \y)}[\lambda \text{CE}(\y, q_{\bphi}(\y|\tilde{\boldsymbol{b}})) + \notag\\
    &\qquad \qquad \qquad \qquad(1-\lambda)\text{KL}(p_{\boldsymbol{T}}(\y|\x) \| q_{\bphi}(\y|\tilde{\boldsymbol{b}}))].
\end{align}
where $\lambda$ balances the contributions of the ground-truth labels and the teacher's soft predictions, and $q_{\boldsymbol{\phi}}(\boldsymbol{y}|\tilde{\boldsymbol{b}})$ denotes the predictive distribution of the server-side FPNN given the received binary features $\tilde{\boldsymbol{b}}$.

From an information-theoretic perspective, minimizing the cross-entropy objective fundamentally drives the system to extract features that are maximally informative about the downstream inference task. Specifically, utilizing the variational distribution $q_{\boldsymbol{\phi}}(\boldsymbol{y}|\tilde{\boldsymbol{b}})$ to approximate the intractable true posterior mathematically maximizes the variational lower bound of the mutual information $I(Y; \tilde{\boldsymbol{b}})$:
\begin{align}
    I(Y; \tilde{\boldsymbol{b}}) \geq \mathbb{E}_{p(\y,\x)p(\tilde{\boldsymbol{b}}|\x)}[\log q_{\bphi}(\y|\tilde{\boldsymbol{b}})].\label{eq: variational}
\end{align}
Crucially, this optimization framework intrinsically aligns with the information bottleneck (IB) principle, which is the essential core of task-oriented communication systems. The standard IB principle dictates maximizing task-relevant information $I(Y; \tilde{\boldsymbol{b}})$ while strictly constraining the total extracted information $I(X; \tilde{\boldsymbol{b}})$ to avoid transmitting task-irrelevant redundancy. Although our loss function $\mathcal{L}$ does not require an explicit regularization term to penalize $I(X; \tilde{\boldsymbol{b}})$, this information constraint is implicitly enforced by the architectural and physical design of our framework.

Unlike continuous real-valued representations, the entropy of the discrete binary features $\tilde{\boldsymbol{b}} \in \{-1, +1 \}^{K_b}$ establishes a hard upper bound on the mutual information, such that $I(X; \tilde{\boldsymbol{b}}) \leq H(\tilde{\boldsymbol{b}}) \leq K_b$. Furthermore, the stochastic bit corruption introduced by the frequency-selective multipath channel, coupled with the selective nature of our subcarrier-based feature calibration, acts as an additional severe physical bottleneck. As illustrated in Fig. \ref{fig:bottleneck}, these compounding structural limitations naturally restrict $I(X; \tilde{\boldsymbol{b}})$. Therefore, training our hybrid-precision framework under these stringent architectural and wireless constraints inherently fulfills the IB optimization, extracting a maximally compact and highly task-relevant representation for edge inference.

\subsection{Gradient Propagation via STE}
The end-to-end training of the proposed hybrid precision framework under the KD-based loss function $\mathcal{L}$ involves the backpropagation with non-differentiable parts, including weight binarization, binary activation, and the digital communication modules (e.g., QAM modulation). To address this issue, we adopt the straight-through estimator (STE) to ensure reasonable gradients for effective backpropagation. For the weight binarization, the derivative of the sign function using STE can be expressed as       
 \begin{figure}[t]
		\centering
		\subfloat[The approximation for sign function and its derivative]{
			\centering
			\includegraphics[width=0.85\linewidth]{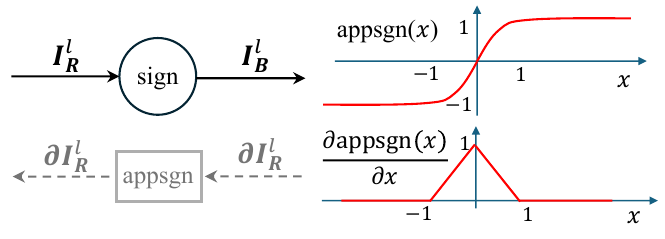}
            \label{subfig:m-1-1}
		}
        
        \subfloat[STE for non-differentiable channel models]{
			\centering
			\includegraphics[width=0.85\linewidth]{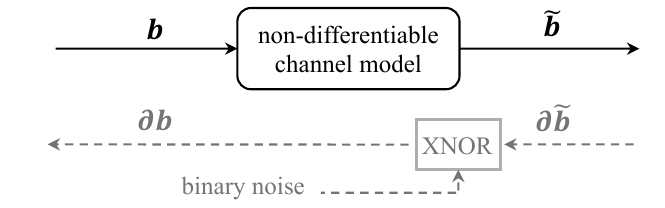}
            \label{subfig:m-1-2}
		}
		\caption{Gradient approximation of sign function and non-differentiable channel models.
		}
        \vspace{-15pt}
		\label{fig:m-1}
\end{figure}
\begin{align}
    \frac{\partial \mathcal{L}}{\partial \mathbf{W_{\rm{R}}}} &= \frac{\partial \mathcal{L}}{\partial\,\text{sign}(\mathbf{W_{\rm{R}}})} \frac{\partial\, \text{sign}(\mathbf{W_{\rm{R}}})}{\partial\mathbf{W_{\rm{R}}}} \overset{\text{STE}}{\approx} \frac{\partial \mathcal{L}}{\partial\,\text{sign}(\mathbf{W_{\rm{R}}})}.
\end{align}

For the binary activation, using STE fails to learn weights near the borders of $-1$ and $+1$, which greatly harms the updating ability of back propagation and restricts the expressiveness of the forward features. To address this, a polynomial step function is used to approximate the forward sign function
\begin{align}
    \text{appsgn}(x) & = \left\{ \begin{array}{lr}
         +1, & x > 1 \\
         2x-x^2, & 1\geq x >0\\
         2x+x^2, & 0\geq x >-1\\
         -1, &  -1 \geq x
    \end{array}
    \right..
\end{align}
Therefore, the derivative of the input features can be expressed as,
\begin{align}
    \frac{\partial \mathcal{L}}{\partial\, \mathbf{I_{\rm{R}}}} & = \frac{\partial\mathcal{L}}{\partial\, \text{sign}(\mathbf{I_{\rm{R}}})} \frac{\partial\, \text{sign}(\mathbf{I_{\rm{R}}})}{\partial \mathbf{I_{\rm{R}}}}\\
    & \overset{\text{STE}}{\approx} \frac{\partial\mathcal{L}}{\partial\, \text{sign}(\mathbf{I_{\rm{R}}})} \frac{\partial\, \text{appsgn}(\mathbf{I_{\rm{R}}})}{\partial \mathbf{I_{\rm{R}}}},
\end{align}
Similarly, the derivative of the learnable bias $\beta$ can be expressed as,    
\begin{align}
    \frac{\partial\, \text{sign}(\mathbf{I_{\rm{R}}+\beta})}{\partial \beta} &\overset{\text{STE}}{\approx} 1.
\end{align}
For the non-differentiable channel models and the digital communication modules, STE can be directly applied as Fig.~\ref{fig:m-1}. 

The proposed STE estimators approximate gradients for the non-differentiable sign and transmission functions during backpropagation. This allows the model to be trained end-to-end by integrating IB-based theoretical foundation with KD. Ultimately, this unified strategy ensures robust convergence and maximizes the preservation of task-relevant information in the hybrid-precision model under varying channel conditions.
\section{Experiments}\label{sec: exp}

\subsection{Experimental Setup}
\subsubsection{Dataset} The experiments are conducted on the ILSVRC12 ImageNet dataset. ImageNet~\cite{russakovsky2015imagenet} is a large-scale dataset with 1000 classes and 1.2 million training images and 50k validation images. Compared to other small-scale datasets commonly used in edge inference studies, it is more challenging and suitable for recent edge inference application scenarios due to its large scale and great diversity.
\subsubsection{Model Architectures}
We adopt ResNet~\cite{he2016deep} architectures as the backbones of the experiments and employ a group of ResNet architectures from small to large sizes, including ResNet-18, ResNet-34, ResNet-50, and ResNet-101. Then, we split the standard ResNet at the middle point into two parts, where the front half is binarized into the on-device BNN using the proposed method, and the latter half is adopted as the server-based FPNN. We name the on-device BNN and the server-based FPNN as \emph{B-$n$} and \emph{FP-$n$}, respectively, where $n$ is the number of layers. As the size of the on-device BNN is expected to be small to fit a wide spectrum of devices, we only consider B-9 and B-15 (i.e., binarized first half of the ResNet-18 and ResNet-34) in our experiments. The summary of layers, BOPs, and FLOPs of all the considered on-device BNNs and server-based FPNNs is presented in Table~\ref{tab:layers-tmp}.

    \begin{table*}
		\caption{The Neural Network Architecture}
        \vspace{-5pt}
		\label{tab:layers-tmp}
        \centering
        \begin{tabular}{lccccccc}
        \toprule
            \textbf{Model Type} & \textbf{Name} & \textbf{BackBone} & \textbf{Layers} & \textbf{BOPs} $(\times 10^9)$& \textbf{FLOPs} $(\times 10^8)$ & \textbf{\#Params} $(\times 10^6)$ & \textbf{Storage} (MB)\\
        \midrule
            \multirow{2}{*}{\textbf{On-Device BNN}}& B-9 & ResNet-18& Conv + [1-bit conv]$\times 8$ & {$0.88$} & {$\mathbf{1.20}$} & $0.68$ &$\mathbf{0.08}$\\
            &{B-15} &ResNet-34& Conv + [1-bit conv]$\times 14$& {$1.57$} & $\mathbf{1.22}$ & $1.34$ & $\mathbf{0.16}$\\
        \midrule
           \multirow{4}{*}{\textbf{Server-Based FPNN}}& {FP-9} & ResNet-18& ResBlock$\times 4$ + Dense & {$0.00$} & {$8.25$} & $11.01$ & $42.01$\\
            &{FP-19} &ResNet-34& ResBlock$\times 9$ + Dense & {$0.00$} & {$19.85$} &$20.43$ & $77.97$\\
            &{FP-28} & ResNet-50&Bottleneck$\times 9$ + Dense & {$0.00$} & {$22.74$} &$24.11$ & $91.99$\\
            &{FP-79} & ResNet-101&Bottleneck$\times 26$ + Dense & {$0.00$} & {$59.80$} &$43.11$ & $164.45$\\
        \bottomrule
        \end{tabular}
        \vspace{-10pt}
	\end{table*}

\subsubsection{Baselines} We compare the proposed hybrid framework with other categories of edge inference. As compared in Table~\ref{tab: intro}, the server-based FPNN, FPNN-Split-Compression, FPNN-Split-VQ, and Full BNN are adopted in the experiments. For the FPNN-Split category, as it exhibits prohibitive communication overheads due to the data amplification effect, we only consider its enhanced versions with additional compression modules and VQ modules, FPNN-Split-Compression and FPNN-Split-VQ. 
\begin{itemize}
    \item \textbf{Server-Based FPNN:} This baseline is to transmit the local data to the edge server, and the FPNN (e.g., ResNet-50) performs inference based on the received local data. 
    \item \textbf{FPNN-Split-Compression:} Following the previous works~\cite{shao2020bottlenet++, shao2021learning}, the bottleneck modules are introduced in these benchmarks to compress the dimensionality of the intermediate features by reducing the channel number from $256$ to $4$, $5$, and $6$. For ease of fair comparison, the floating-point value is formatted with double-precision using 64 bits. 
    \item \textbf{FPNN-Split-VQ:} Following the previous works~\cite{xie2023robust}, vector quantization modules are introduced to make the floating-point features into the discrete representations. By adopting the codebook with 256 codewords, each $7$-dimensional feature vector is quantized. 
    \item \textbf{Full BNN:} To comprehensively compare the performance of the proposed hybrid framework and BNNs, we select representative BNNs with different binarization techniques, as summarized in Section~\ref{subsec:related-BNN}. They include XNOR-Net~\cite{rastegari2016xnor}, ABC-Net~\cite{lin2017towards}, Bi-Real-Net~\cite{liu2018bi}, IR-Net~\cite{qin2020forward}, BNAS~\cite{ding2021bnas}, NASB~\cite{zhu2020nasb}, Si-BNN~\cite{wang2020sparsity}, ProxyBNN~\cite{he2020proxybnn}, RBNN~\cite{qiu2022rbnn}, ReActNet~\cite{liu2020reactnet}, ReCU~\cite{xu2021recu}, Bi-half~\cite{li2022equal}, SiMaN~\cite{lin2022siman}, AdaBin~\cite{tu2022adabin}, and DIR-Net~\cite{qin2023distribution}.
\end{itemize}

\subsubsection{Implementation Details}
The considered on-device BNNs encode the image from the ImageNet dataset and output the binary features $\mathbf{I_B}$ with the size $\mathbf{I_B}\in \{-1, 1 \}^{256\times 14\times 14}$ with the scaling factors $\boldsymbol{\Upsilon} \in \mathbb{R}^{14\times 14}$. Each floating-point value in $\boldsymbol{\Upsilon}$ is quantized into int8 format and represented by 8 bits. Then, the bit sequence will be modulated by QAM-16 modulation into complex signals. We simulate the multipath fading channel with $\Psi=8$ signal paths and the delay spread $\gamma=4$. For the OFDM transmission, each OFDM symbol has $N_{\text{FFT}} = 256$ subcarriers and CP length is set to $N_{\text{CP}}= 16$, ensuring it exceeds the delay spread of the multipath channel. For each image, the binary features $\mathbf{I_B}$ take $49$ OFDM symbols and the binarized  $\boldsymbol{\Upsilon}$ take $2$ OFDM symbols. Therefore, $51$ OFDM symbols are transmitted for the proposed hybrid framework. For the KD-based training of the proposed framework, a full-precision ResNet-101 serves as the teacher for the ResNet-101-based hybrid model, while a full-precision ResNet-50 is employed as the teacher for all other configurations. The inference performance is evaluated by the top-1 accuracy. The binary and floating-point computation overhead are evaluated by BOP and FLOP. Note that $\text{OP} =\text{BOP} = \text{FLOP}/64$ if we comprehensively consider the computational overhead of BOP and FLOPs into a single metric OPs.

\subsection{Model Capacity of Hybrid Edge Inference}
\begin{figure}[t]
		\centering
		\includegraphics[width=8cm]{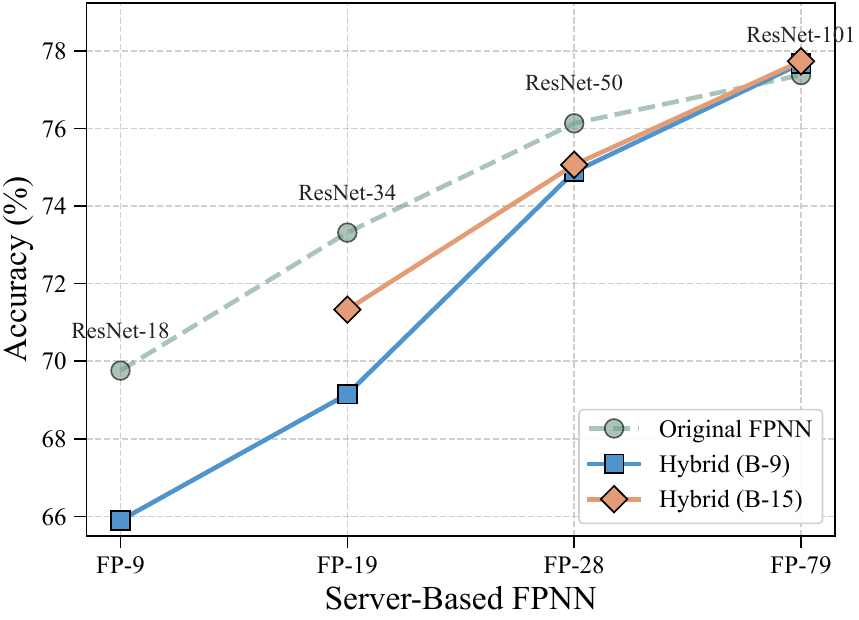}
        \vspace{-5pt}
		\caption{The accuracy of the hybrid framework with different pairs of on-device BNNs and server-based FPNNs, and the corresponding original FPNNs. 
		}\label{fig:exp-capacity}
        \vspace{-10pt}
\end{figure}

\begin{table}[t]
		\caption{Performance comparison of full BNNs and the proposed hybrid framework based on ResNet-18 backbone}
        \vspace{-5pt}
		\label{tab:comparison-with-BNNs-1}
        \centering
        \begin{tabular}{lccc}
        \toprule
         Methods & Acc $(\%)$ & BOPs $(\times 10^9)$ & FLOPs $(\times 10^8)$ \\
        \midrule
        XNOR-Net      & $51.20$ &  $1.70$  &  $1.41$  \\
        ABC-Net       & $61.00$  & $5.10$   & $4.89$    \\
        Bi-Real-Net   & $56.40$ & $1.68$ & $1.63$  \\
        IR-Net        & $58.10$  & $1.68$ & $1.63$ \\
        BNAS          & $58.76$  & $1.68$ & $1.63$  \\
        NASB          & $60.50$  & $1.68$ & $1.63$  \\
        Si-BNN        & $58.90$  &$1.70$  &  $1.63$  \\
        ProxyBNN      & $63.70$  & $1.68$  &  $1.63$  \\
        RBNN          & $59.90$  &  $1.68$  &  $1.63$\\
        ReActNet      & $65.50$  &  $1.68$  &  $1.63$ \\
        ReCU          & $61.00$  & $1.68$  &  $1.63$ \\
        Bi-half       & $60.40$  &$1.68$ & $1.63$ \\
        SiMaN         & $60.10$  &$1.68$ & $1.63$\\
        AdaBin        & $63.10$  &$1.68$ & $1.63$\\
        DIR-Net       & $60.40$  &$1.68$ & $1.63$\\
        \textbf{Hybrid (B-9)}& $\boldsymbol{65.90}$  & $\boldsymbol{0.88}$ & $\boldsymbol{1.20}$ \\
        \bottomrule
        \end{tabular}
        \vspace{-10pt}
	\end{table}

    \begin{table}[t]
		\caption{Performance comparison of full BNNs and the proposed hybrid framework based on ResNet-34 Backbone}
        \vspace{-5pt}
		\label{tab:comparison-with-BNNs-2}
        \centering
        \begin{tabular}{lccc}
        \toprule
         Methods& Acc $(\%)$ & BOPs $(\times 10^9)$ & FLOPs $(\times 10^8)$\\
        \midrule
        XNOR-Net      &  $56.49$ & $3.53$ &  $2.50$  \\
        ABC-Net       &  $66.70$ & $10.59$& $2.80$    \\
        Bi-Real-Net   &  $62.20$ &$3.53$&$2.50$\\
        IR-Net        &  $62.90$ &$3.53$&$2.50$\\
        BNAS          &  $59.81$ &$3.53$&$2.50$\\
        NASB          &  $64.00$ &$3.53$&$2.50$\\
        Si-BNN        &  $63.30$ &$-$&$-$\\
        ProxyBNN      &  $66.30$ & $3.53$   &  $2.50$\\
        RBNN          &  $63.10$ & $3.53$   &  $2.50$ \\
        ReCU          &  $65.10$ &    $3.53$   &  $2.50$  \\
        Bi-half       &  $64.17$ & $3.53$&$2.50$ \\
        SiMaN         &  $63.90$ & $3.53$&$2.50$ \\
        AdaBin        &  $66.40$ & $3.53$   &  $2.50$   \\
        DIR-Net       &  $64.10$ &$3.53$  &  $2.50$  \\
        \textbf{Hybrid (B-9)}&   ${69.15}$ & $\boldsymbol{0.88}$   & $\boldsymbol{1.20}$    \\
        \textbf{Hybrid (B-15)} &  $\boldsymbol{71.33}$ & $1.57$ & $1.22$\\
        \bottomrule
        \end{tabular}
        \vspace{-10pt}
	\end{table}
Before comparing the proposed hybrid framework with co-inference frameworks and on-device BNNs, we first investigate the model capacity and expressivity of the hybrid-precision design by comparing it with the original FPNN. We train hybrid frameworks with on-device BNNs, B-9 and B-15, with the server-based FPNNs including FP-9, FP-19, FP-28, and FP-79. Classification accuracy of all the evaluated methods is presented in Fig.~\ref{fig:exp-capacity}. The original FPNN is expected to be the performance upper bound, showing the performance gap between the hybrid model and the original FPNN model. This performance gap is most evident at the FP-9 level, where around $4\%$ accuracy gap can be observed. And the accuracy of all methods converges as the server-side model scales toward FP-79 (ResNet-101). The initial 4\% performance gap between Hybrid (B-9) and the original FPNN at the lower tier narrows to a negligible margin at the highest tier. This suggests that a powerful server-side model can refine coarse and binarized features to a level nearly identical to full-precision performance. Furthermore, since the on-device BNNs only utilize the front halves of ResNet-18 and ResNet-34, configurations such as B-9/FP-79 and B-15/FP-79 have \emph{fewer} total layers than their corresponding full-precision counterparts, ResNet-50 and ResNet-101. This highlights the superior balance of efficiency and accuracy achieved by the proposed hybrid solution.
\subsection{Performance Comparison: Hybrid Framework vs. Full BNNs}
We compare the proposed hybrid frameworks against state-of-the-art full BNNs deployed solely on the edge device. For the hybrid framework, binary features are transmitted over multipath fading channels with an SNR $=20$dB to the server-based FPNN. In contrast, full BNNs execute the entire inference process locally on the edge device. However, the existing BNNs research is mainly based on lightweight backbones such as ResNet-18 and ResNet-34. To ensure the fair comparison, we only compare under the backbones (i.e., ResNet-18 and ResNet-34) of the existing BNNs considered. 

In Table~\ref{tab:comparison-with-BNNs-1} and~\ref{tab:comparison-with-BNNs-2}, the experimental results demonstrate that the proposed hybrid-precision framework achieves superior accuracy while maintaining significantly lower on-device computational costs. For the ResNet-18 configuration, Hybrid (B-9) reaches 65.90\% accuracy, surpassing leading binarized models such as ReActNet and ProxyBNN. Notably, the hybrid approach achieves this while requiring only $0.88 \times 10^9$ BOPs and $1.20 \times 10^8$ FLOPs, which is nearly half the computational workload of standard BNNs. This performance advantage extends to the ResNet-34 backbone, where Hybrid (B-15) achieves 71.33\% accuracy, outperforming the best-performing traditional BNN, ABC-Net, by a substantial margin of 4.63\%. 
Furthermore, while traditional BNNs are often incompatible with large-scale backbones like ResNet-101 due to the prohibitive local memory and computational requirements, the proposed hybrid framework overcomes these limitations. Distributing the workload, it enables resource-constrained edge devices to leverage the power of deep architectures.

 \begin{figure*}
		\centering
		\subfloat[ResNet-18 Backbone]{
			\centering
			\includegraphics[width=0.29\linewidth]{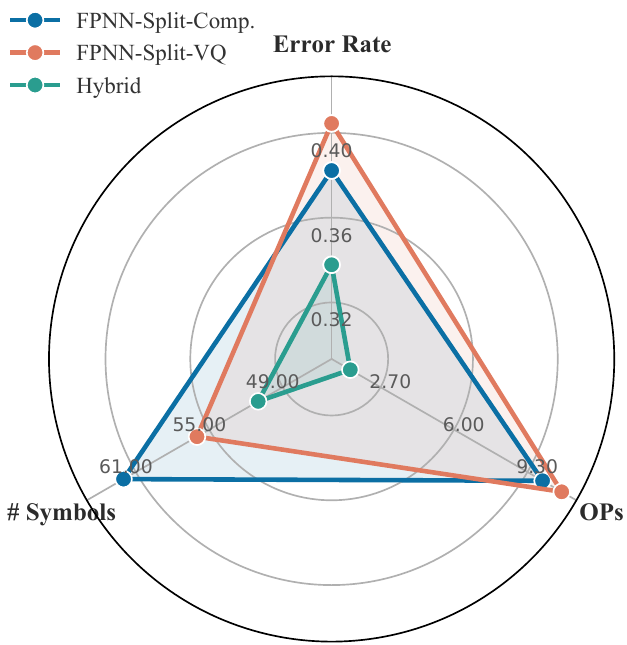}
		}
        \subfloat[ResNet-34 Backbone]{
			\centering
			\includegraphics[width=0.29\linewidth]{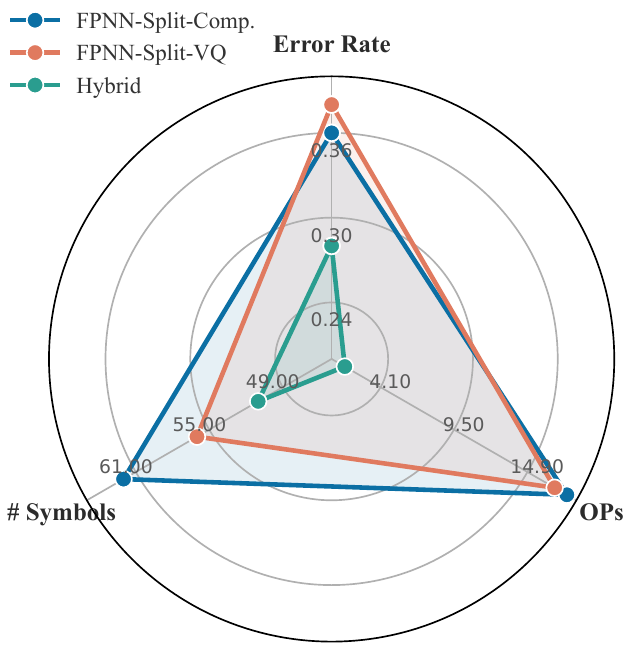}
		}
		\subfloat[ResNet-101 Backbone]{
			\centering
			\includegraphics[width=0.29\linewidth]{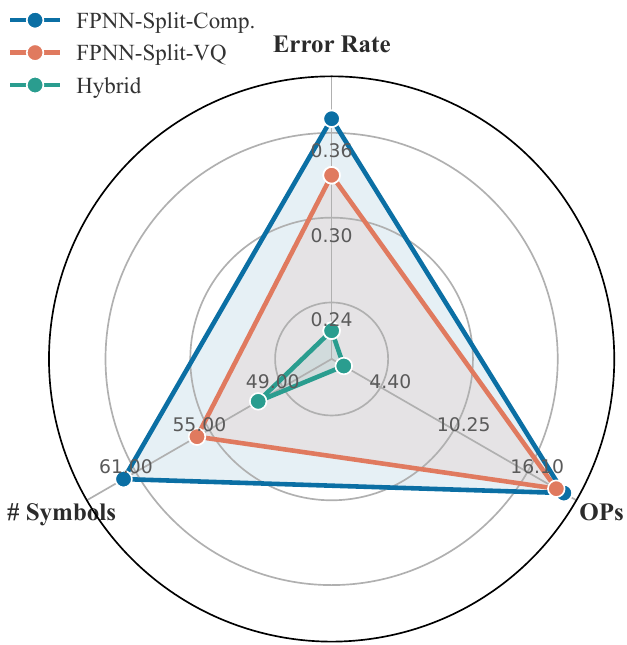}
		}
		\caption{Performance comparison of the proposed hybrid, FPNN-Split-Compression, and FPNN-Split-VQ under SNR=20dB.
		}
        \vspace{-10pt}
		\label{fig:radars}
	\end{figure*}

     \begin{figure*}
		\centering
		\subfloat[ResNet-18 Backbone]{
			\centering
			\includegraphics[width=0.29\linewidth]{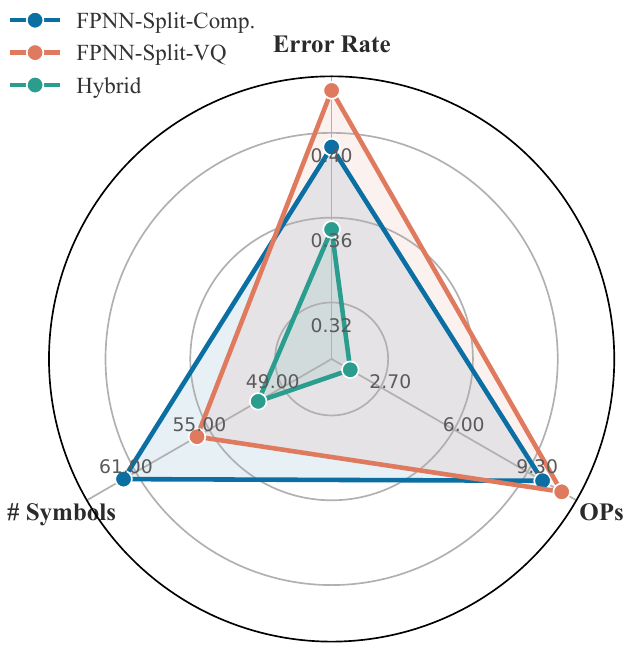}
		}
        \subfloat[ResNet-34 Backbone]{
			\centering
			\includegraphics[width=0.29\linewidth]{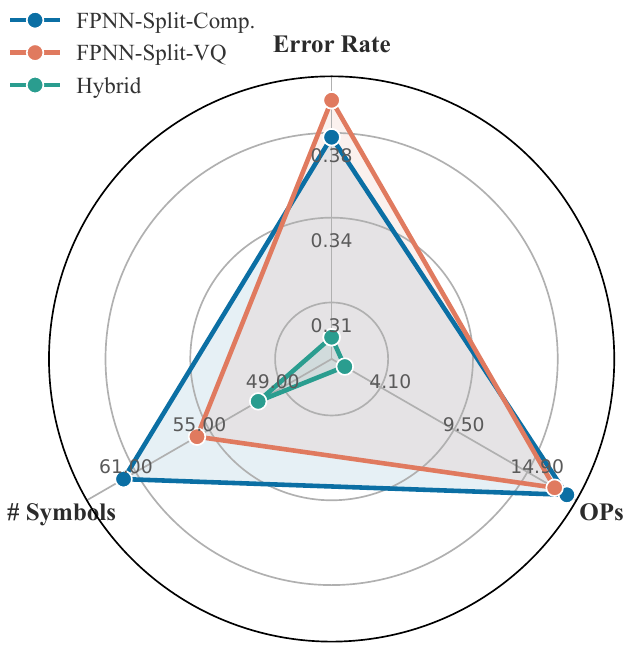}
		}
		\subfloat[ResNet-101 Backbone]{
			\centering
			\includegraphics[width=0.29\linewidth]{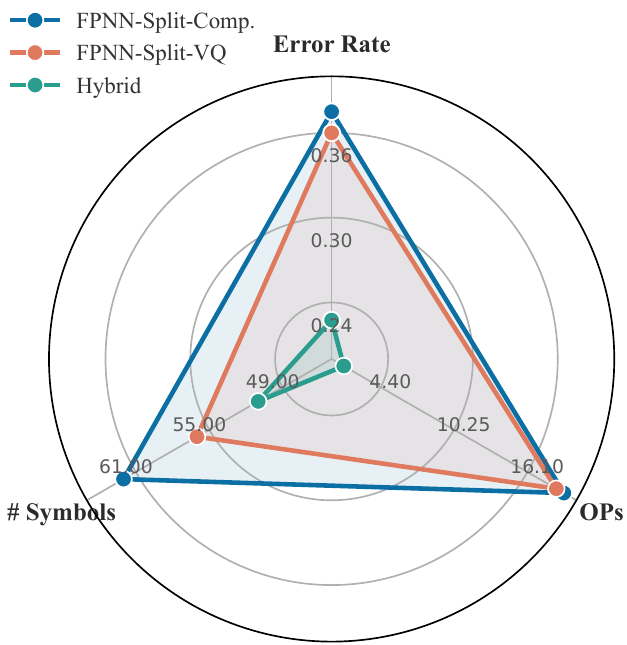}
		}
		\caption{Performance comparison of the proposed hybrid, FPNN-Split-Compression, and FPNN-Split-VQ under SNR$=10$dB.
		}
        \vspace{-10pt}
		\label{fig:radars-2}
	\end{figure*}
    
\subsection{Performance Comparison: Hybrid Framework vs. Edge Co-Inference Frameworks}
Then, we compare the proposed hybrid design against two established edge co-inference baselines: FPNN-Split-Comp. and FPNN-Split-VQ. The comparison is conducted under multipath fading channels at two SNR levels: 20dB and 10dB. We analyze the frameworks across three critical performance dimensions: communication overhead, computational complexity, and classification error rate.

In Fig.~\ref{fig:radars}, the experimental results at SNR $=$ 20dB demonstrate that the hybrid framework significantly outperforms both splitting-based solutions across all backbone architectures. The hybrid model consistently occupies the innermost region of the radar charts, showing the lowest error rate, the fewest required OFDM symbols, and drastically reduced computational OPs. For example, in the ResNet-101 configuration, the hybrid framework achieves an error rate of even less than 0.24, whereas the FPNN-Split-VQ and FPNN-Split-Comp. suffer from higher error rates and nearly triple the computational cost. This highlights the hybrid framework's superior ability to extract and transmit compact, noise-resilient features. When the channel conditions degrade to an SNR of 10dB, as shown in Fig.~\ref{fig:radars-2}, all models experience a decline in classification accuracy due to increased signal interference. However, the performance hierarchy remains unchanged: the hybrid framework continues to provide the best balance of resource efficiency and accuracy, proving its robustness in more challenging wireless environments despite the lower absolute accuracy.

\subsection{Robustness}
\begin{figure}
		\centering
		\subfloat[Hybrid Precision (B-9, FP-9)]{
			\centering
			\includegraphics[width=0.85\linewidth]{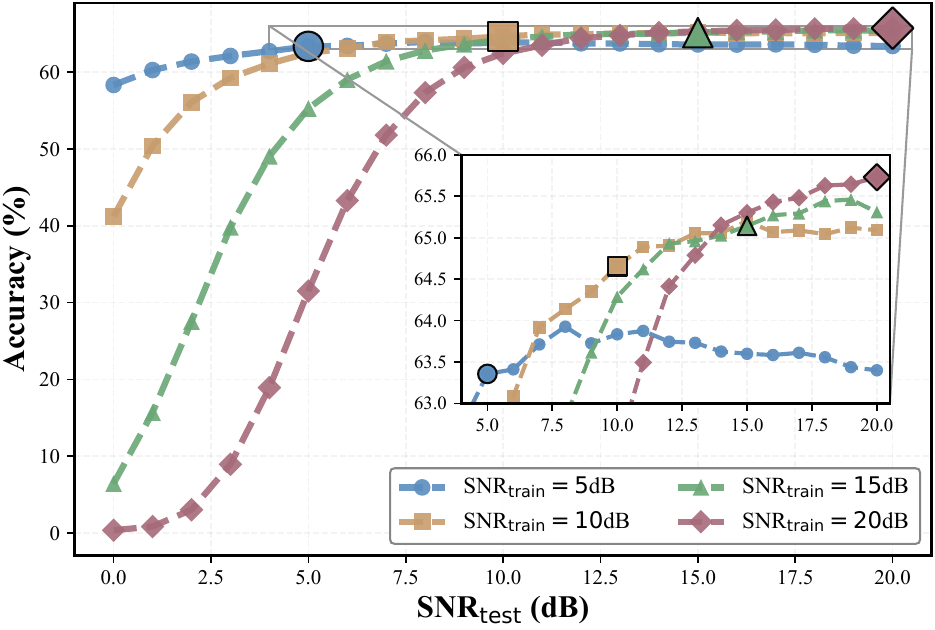}
		}
        \vspace{-5pt}
        
        \subfloat[Hybrid Precision (B-9, FP-79)]{
			\centering
			\includegraphics[width=0.85\linewidth]{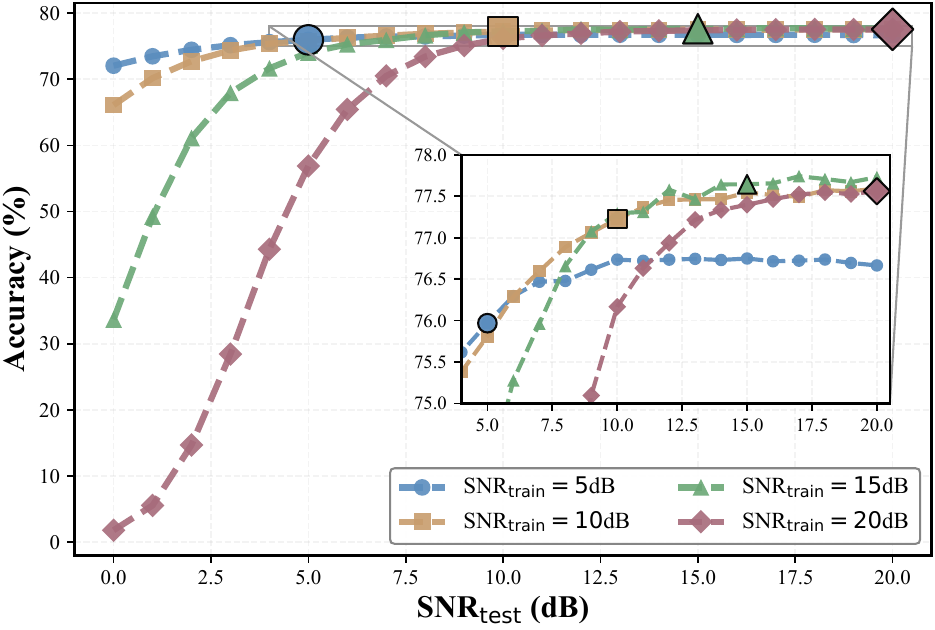}
		}
		\caption{Performance of the hybrid precision framework under varying testing SNR.
		}
        \vspace{-15pt}
		\label{fig:robust}
\end{figure}
To evaluate the robustness of the proposed hybrid framework, we analyze the classification accuracy under varying testing SNR conditions ($\rm{SNR}_{\rm{test}}$) for models trained at noise levels $\rm{SNR}_{\rm{train}} = 5, 10, 15, \text{and } 20$ dB. As illustrated in Fig.~\ref{fig:robust}, there is a clear trade-off between accuracy and noise resilience based on the training environment. Models trained at high SNRs (e.g., 20 dB) achieve the highest peak accuracy when testing conditions are ideal but suffer from a cliff effect, where performance drops to near zero as $\rm{SNR}_{\rm{test}}$ decreases below 5 dB. In contrast, models trained under noisier conditions (e.g., 5 dB) exhibit significantly higher robustness; for the Hybrid (B-9, FP-9) model, training at 5 dB maintains approximately 60\% accuracy even at $0$ dB $\rm{SNR}_{\rm{test}}$, whereas the 20 dB trained model fails completely. This suggests that low-SNR training effectively acts as a noise-regularization technique, allowing the server-based FPNN to learn to decode heavily distorted binary features. The impact of the server-based backbone complexity is also evident when comparing the FP-9 and FP-79 configurations. While both frameworks follow identical robustness trends, the hybrid (B-9, FP-79) system provides a substantially higher performance ceiling, reaching a peak accuracy of approximately 78\% compared to the 66\% achieved by the FP-9 variant. Notably, the larger server model also demonstrates superior baseline robustness; when trained at 5 dB, it maintains over 72\% accuracy at $0$ dB $\rm{SNR}_{\rm{test}}$. This indicates that a more powerful server-side backbone not only improves absolute accuracy but also possesses a stronger capacity to reconstruct information from noisy edge-transmitted features. Consequently, for deployment in highly volatile wireless environments, a hybrid configuration utilizing a low $\rm{SNR}_{\rm{train}}$ and a high-capacity server backbone offers the most reliable balance of efficiency and all-terrain performance.
\begin{figure}
		\centering
		\subfloat[$\rm{SNR}_{\rm{\rm{train}}}=5 \rm{dB}$]{
			\centering
			\includegraphics[width=0.85\linewidth]{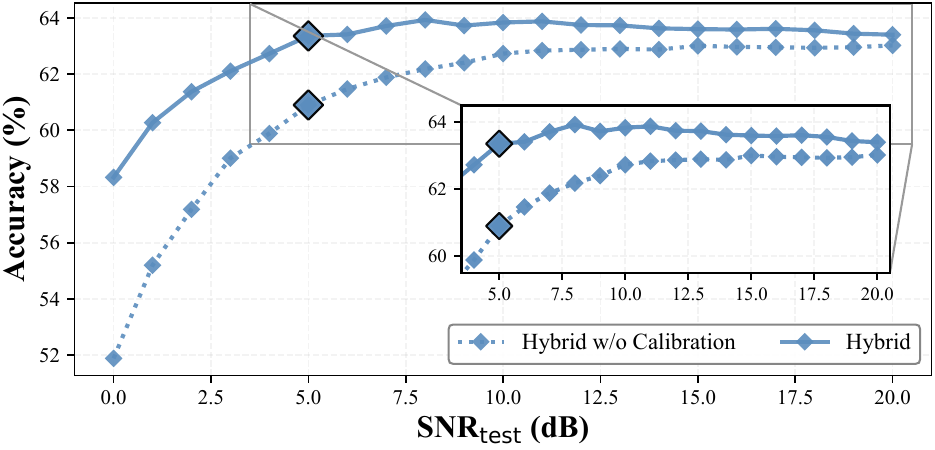}
		} 
        \vspace{-5pt}
        
        \subfloat[$\rm{SNR}_{\rm{\rm{train}}}=10 \rm{dB}$]{
			\centering
			\includegraphics[width=0.85\linewidth]{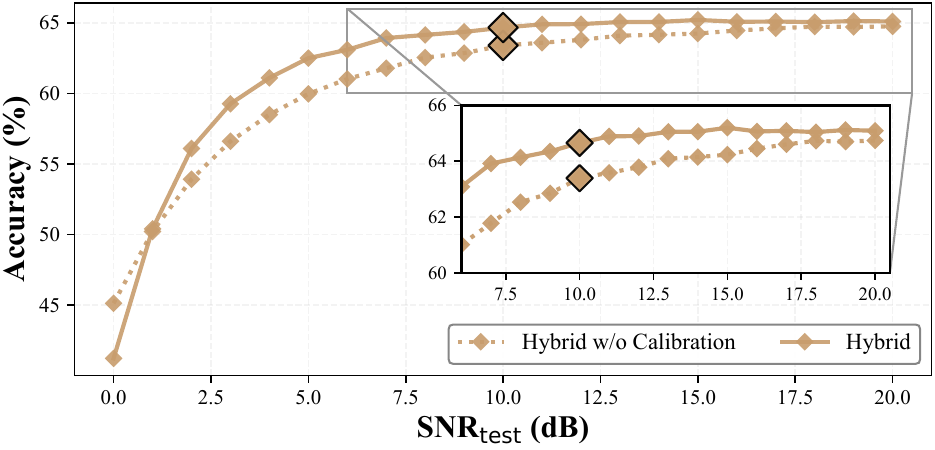}
		}
        \vspace{-5pt}

        \subfloat[$\rm{SNR}_{\rm{\rm{train}}}=15 \rm{dB}$]{
			\centering
			\includegraphics[width=0.85\linewidth]{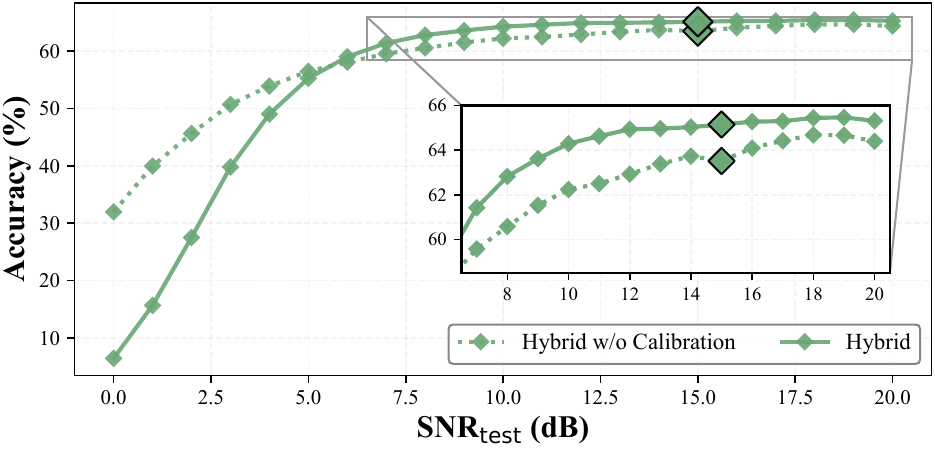}
		}
        \vspace{-5pt}
        
        \subfloat[$\rm{SNR}_{\rm{\rm{train}}}=20 \rm{dB}$]{
			\centering
			\includegraphics[width=0.85\linewidth]{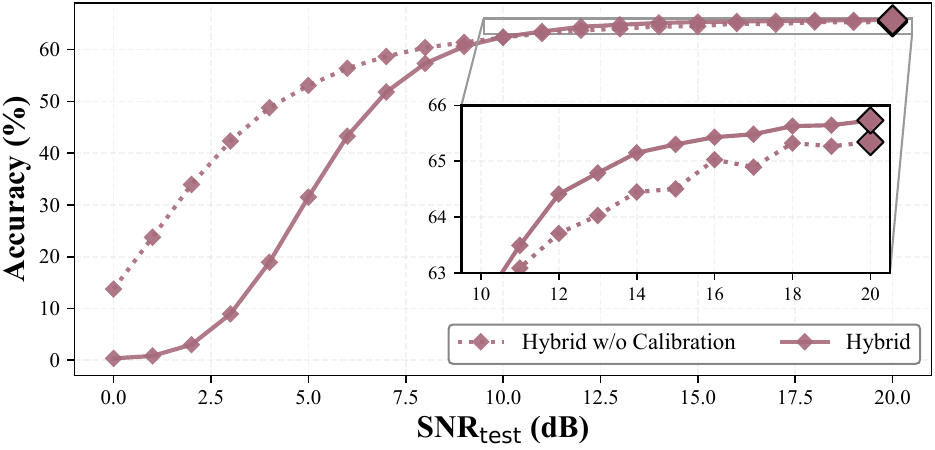}
		}
        \vspace{-5pt}
		\caption{Performance comparison on subcarrier-based feature calibration under varying testing SNR 
		}
        \vspace{-15pt}
		\label{fig:ablation}
\end{figure} 
To evaluate the subcarrier-based feature calibration module, we conducted an ablation study comparing the standard hybrid precision framework against a version without calibration named Hybrid w/o Calibration. As illustrated in Fig.~\ref{fig:ablation}, the calibration module consistently enhances classification accuracy and robustness within reasonable operational regions by effectively mitigating binary feature distortion caused by multipath fading. This specialization introduces an unavoidable trade-off resulting in increased sensitivity and sharper performance declines when testing SNRs fall significantly below training thresholds. By enabling the server-side FPNN to recover high-fidelity information from noise-corrupted streams, the calibration mechanism ensures superior accuracy across the most critical operating environments. To achieve full adaptivity in various channel conditions, recent techniques such as attention modules~\cite{xu2021wireless} and Hypernetworks~\cite{10621322} can be integrated into the proposed hybrid-precision framework.

\section{Conclusion}\label{sec: conc}
This paper investigated the design of a hybrid-precision task-oriented communication that integrates efficient architectural binarization with edge co-inference. By deploying a binarized front-end on resource-constrained edge devices and a full-precision back-end on the edge server, the proposed framework successfully addresses the limitations of traditional model splitting and full BNN implementations. The inclusion of subcarrier-based calibration and binary feature recovery ensures robust performance over challenging multipath fading wireless environments. Combined with a KD-based optimization strategy, this hybrid approach effectively bridges the utility gap caused by binarization while maintaining minimal on-device computational overhead. Experimental results on the large-scale ImageNet dataset validate the effectiveness of the proposed system.

Future research will extend this hybrid-precision paradigm to multi-user collaborative scenarios, exploring distributed feature extraction and subcarrier allocation for large-scale edge networks. Furthermore, we aim to adapt this framework to multi-modal large language models (LLMs) for maintaining cross-modal semantic alignment while significantly reducing the memory and communication overhead of generative AI tasks at the wireless edge.

\appendices

\bibliographystyle{IEEEtran}
\bibliography{IEEEabrv,ref}

\end{document}